\documentclass[10pt,journal,compsoc]{IEEEtran}

\usepackage{amssymb}
\usepackage[nocompress]{cite}
\usepackage[pdftex]{graphicx}
\graphicspath{{diagrams/}}
\DeclareGraphicsExtensions{.pdf}
\usepackage{makecell}
\usepackage{url}

\usepackage{hyperref}
\usepackage{enumitem}

\newcommand{\oecode}[1]{\textbf{#1}}
\newcommand{\response}[1]{``\emph{#1}''}
\newcommand{\rid}[1]{[#1]}
\newcommand{\eg}{\textit{e}.\textit{g}., }
\newcommand{\ie}{\textit{i}.\textit{e}., }
\newcommand{\etal}{\textit{et al}.}

\IEEEpubid{
\begin{minipage}{\textwidth}\ \\[12pt]
\copyright\ 2022 EU Copyright. Personal use of this material is permitted.
Permission from IEEE must be obtained for all other uses, in any current or future media,
including reprinting/republishing this material for advertising or promotional purposes,
creating new collective works, for resale or redistribution to servers or lists, or
reuse of any copyrighted component of this work in other works.
DOI: \href{https://doi.org/10.1109/TSE.2022.3208210}{10.1109/TSE.2022.3208210}
\end{minipage}
}

\begin{document}

\title{Impact of Software Engineering Research in Practice:
A Patent and Author Survey Analysis}

\author{
	Zoe~Kotti,
	Georgios~Gousios,
	and Diomidis~Spinellis,~\IEEEmembership{Senior Member,~IEEE}
	\IEEEcompsocitemizethanks{
		\IEEEcompsocthanksitem D. Spinellis and G. Gousios are with the
			Department of Software Technology,
			Delft University of Technology,
			The Netherlands.\protect\\ 
			E-mail: \{D.Spinellis,G.Gousios\}@tudelft.nl
		\IEEEcompsocthanksitem D. Spinellis and Z. Kotti are with the
			Department of Management Science and Technology,
			Athens University of Economics and Business,
			Greece.\protect\\
			E-mail: \{dds,zoekotti\}@aueb.gr
	}
}

\markboth{}%
{Kotti \MakeLowercase{\etal}: Impact of Software Engineering Research in Practice:
A Patent and Author Survey Analysis}

\IEEEtitleabstractindextext{%
\begin{abstract}
Existing work on the practical impact of software engineering (SE) research
examines industrial relevance rather than adoption of study results,
hence the question of how results have been practically applied remains open.
To answer this and investigate the outcomes of impactful research,
we performed a quantitative and qualitative analysis
of 4\,354 SE patents citing 1\,690 SE papers published in four leading SE venues
between 1975--2017.
Moreover,
we conducted a survey on 475 authors of 593 top-cited and awarded publications,
achieving 26\% response rate.
Overall,
researchers have equipped practitioners with various tools, processes, and methods,
and improved many existing products.
SE practice values knowledge-seeking research and is impacted
by diverse cross-disciplinary SE areas.
Practitioner-oriented publication venues appear more impactful than researcher-oriented ones,
while industry-related tracks in conferences could enhance their impact.
Some research works did not reach a wide footprint due to limited funding resources or
unfavorable cost-benefit trade-off of the proposed solutions.
The need for higher SE research funding could be corroborated
through a dedicated empirical study.
In general,
the assessment of impact is subject to its definition.
Therefore, academia and industry could jointly agree on a formal description
to set a common ground for subsequent research on the topic.
\end{abstract}

}

\maketitle

\IEEEraisesectionheading{\section{Introduction}\label{sec:introduction}}
\IEEEPARstart{I}{n} 2018,
the field of software engineering (SE) marked the 50th anniversary
of its first two-year conference series---the
1968--69 NATO Conferences on Software Engineering~\cite{NR69,BR70}.
Despite its relatively short period of existence,
a lot of research has been performed in SE
during these 50 years,
composing a large body of information.
In the meantime,
numerous software and technology-related companies have emerged,
partially as a result of hardware advancement and cloud computing~\cite{MG11},
forming a multi-trillion dollar industry~\cite{Sta18}.
This growth both in terms of knowledge and market share raises the question of
how these two relate,
and to what extent research may have impacted industry.
In this context,
we define as \emph{impact} \textbf{the direct or indirect incorporation
of a software engineering study's output
in an industrial setting,
for example, in an industrial software development tool, process,
marketable product, or service}.

In the scope of this study,
we consider \emph{software engineering} the discipline
that systematically employs computer science knowledge and principles
to develop new methods and tools to improve software development.
The discipline's areas include
software requirements,
design,
construction,
testing,
maintenance,
configuration management,
quality,
SE management,
SE models and methods, and
SE process~\cite{swebok}.
The application process is based on systematic, disciplined, and
quantifiable SE approaches, and
is influenced by cross-disciplinary areas, namely,
mathematics,
general management,
project management, and
systems engineering~\cite{swebok}.
Note that our definition distinguishes foundational computer science research
(\eg devising a new static analysis method,
a test prioritization algorithm, or
a requirements definition language)
from that performed in SE contexts.
For the described examples to be considered SE research,
we expect them to be accompanied with empirical evaluation through,
for example, repository mining, a developer survey, or a case study.

Existing work on the practical impact of SE research
examines industrial relevance rather than adoption of study results.
A variety of interviews and literature reviews have been conducted,
mainly in domain-specific contexts such as the ACM SIGSOFT Impact Project~\cite{OCEK00},
to assess the relation of research to industrial needs,
highlight gaps between the two, and
suggest best practices for collaborative projects.
However,
the question of how research results have been practically applied remains open.

To tackle this question and investigate the outcomes of impactful SE research,
we performed a quantitative and qualitative analysis
of SE patents citing SE research from four leading SE venues.
Patents are by definition practical applications of technology,
and are frequently employed as an estimator of the academic research impact
(\eg in the works by Narin \etal~\cite{NHO97},
Estublier \etal~\cite{ELHC05}, and
the National Academy of Engineering~\cite{Nat03}).
Software patents have increased rapidly in number,
comprising 15\% of all patents~\cite{BH07}.
Most of them are acquired by large manufacturing firms
from the computers, electronics, and machinery industries~\cite{BH07}.
Furthermore,
we conducted a survey on authors
of highly recognized SE publications
to examine impactful types, areas, methods, and outcomes of SE research
as well as their footprint on information technology, society, and industry.

Our findings suggest that SE researchers have equipped practitioners
with various tools, processes, and methods,
and improved many existing products.
SE practice seems to value knowledge-seeking research and is impacted
by diverse cross-disciplinary SE areas.
Practitioner-oriented publication venues appear more impactful than researcher-oriented ones,
while industry-related tracks in conferences could enhance their impact.
Two main obstacles to research adoption seem to be insufficient funding
and the unfavorable cost-benefit trade-off of the produced solutions.
The study's contributions are:
\begin{itemize}[leftmargin=*]
\item the systematic collection of top-rated SE research,
\item the systematic collection of SE patents citing SE research,
\item the categorization of research according to its type, methods,
SE area, and industrial application domain, and
\item the synthesis of the preceding results into a taxonomy of the main
practical impacts of SE research.
\end{itemize}

In the following Section we present an overview of existing work
on the practical impact of SE research.
We then describe the research questions and study methods
in Section~\ref{sec:methods}, and
present the research results in Section~\ref{sec:results}.
An extensive discussion of the study findings is included
in Section~\ref{sec:discussion}.
The study is complemented by the associated limitations in Section~\ref{sec:limitations},
followed by our conclusions in Section~\ref{sec:conclusions}.
Based on published guidelines~\cite{IHG12},
the code,\footnote{\url{https://doi.org/10.5281/zenodo.6780414}}
survey questionnaire, anonymized responses, and
produced data\footnote{\url{https://doi.org/10.5281/zenodo.7090818}}
associated with this study are publicly available online,
and can be used for replication or further empirical research.

\section{Related Work}
\label{sec:related}
We analyze and synthesize related work on the practical impact of SE research
according to the study objective (\ie practical impact),
and the two employed method axes (\ie patent and author survey analyses).
For this,
we employ the classification scheme by Lo \etal~\cite{LNZ15},
and classify related work into three areas:
research related to the ACM SIGSOFT Impact Project;
literature reviews and surveys in SE
evaluating the relationship between academia and industry;
and ranking studies assessing the impact of SE researchers,
institutions,
or publication venues.
To cover the patent axis,
we extend this classification scheme
with studies evaluating the impact of SE research based on patent metadata.

\subsection{ACM SIGSOFT Impact Project}
In the early 2000s, the Impact Project~\cite{OCEK00} was established,
in an attempt ``to help both the research and practitioner community
to understand each other better'',
in order to strengthen their cooperation,
and also avail funding agencies to maximize their return on investment
in SE research.
In general,
the project aims to study the impact that SE research has had
upon software development practice.

The project uncovers state-of-the-art software technologies
in specific areas,
and examines their influence by former research work,
through literature searches and personal interviews~\cite{OGKW05,OGKW08}.
Specific areas include
software configuration management~\cite{ELHC05},
modern programming languages~\cite{RSB05},
software testing and analysis~\cite{CR06},
middleware technology~\cite{EAS08},
inspections, reviews, and walkthroughs~\cite{RCJL08}, and
software resource estimation~\cite{BV11}.

According to the area-specific studies,
academic research tools and services
have been adopted by major industrial projects
and have influenced various fields.
Although some original ideas require a long time (up to 15--20 years~\cite{EAS08}),
deep reworking, and re-engineering to apply in industrial practice,
the constant flow of researchers between industry and academia can expedite adoption.
By employing academic research techniques,
laboratories have reported up to 95\% increase in defect detection before testing,
50\% cost reduction for newly developed source code lines, and
up to 50\% shortening of delivery times.

Conversely, companies have contributed to academia
in estimation and mathematical approaches,
advanced project planning,
and flexible and realistic models~\cite{BV11}.
In estimation and mathematics
we distinguish the rise of analogy, expert judgment, hybrid,
and Bayesian approximation methods.
Motivated by corporate projects,
academic ones have started balancing their deadlines and activities better,
and prioritizing features according to schedule, cost, and
quality requirements.
Finally,
SE models have become more adaptable to the diverse programming languages,
new standards,
and techniques such as rapid application development,
and more realistic on the basis of ten evaluation criteria:
\emph{definition},
\emph{accuracy},
\emph{scope},
\emph{objectivity},
\emph{constructiveness},
\emph{detail},
\emph{stability},
\emph{ease of use},
\emph{prospectiveness}, and
\emph{parsimony}~\cite{BV11}.

\subsection{Literature Reviews and Surveys}
\label{sec:reviews}
In 2013,
Misirli \etal~\cite{MCBT13} conducted in-depth interviews with twelve practitioners
who were actively collaborating with them at that time
in three industrial software analytics projects.
These projects involved defect, effort, and quality prediction.
Their aim was to explore practitioners' expectations,
and ways to employ software analytics solutions in policy making.
Respondents suggested enhancing the examined solutions with
defect-severity or defect-type prediction,
defect location,
and phase-- or requirement-level effort estimation.
Furthermore,
they stressed the need for collecting accurate and complete data through the provided solutions,
and integrating these solutions into existing systems
(\eg by combining defect prediction results with test interfaces
to determine which interfaces to test first).

Around the same time,
Beecham \etal~\cite{BORB13} conducted interviews with practitioners
from ten companies of various sizes
to assess the impact
of Global Software Engineering (GSE) research papers in practice.
GSE research is regarded as useful,
and participants argue that studying the topic might improve performance.
Still, none was found to have actually consulted the GSE literature.
Practitioners mostly refer to books, blogs, forums, short reports,
and their past experience to resolve problems in GSE.
Some respondents associated GSE with general project management.
The authors argue that GSE research should be relevant
(\ie reflecting the needs of practice),
documented in short, evidence-based, and readable papers with validated findings,
disseminated more widely as grey literature, and
advertised through social media.

Through another survey on 512 Microsoft practitioners,
Lo \etal~\cite{LNZ15} examined the relevance of SE research to practice.
Participants rated the relevance of 571 ICSE, ESEC/FSE, and FSE papers
published between 2009--2014:
71\% of all ratings were essential or worthwhile,
while no correlation was observed between citation counts and relevance scores.
Reasons behind research ideas rated as ``unwise'' include:
unneeded tools;
non-actionable empirical studies;
generalizability issues;
cost outweighing benefit;
questionable assumptions;
disbelief in the proposed solution;
better alternatives or more crucial problems to handle; and
side effects of the suggested solution.
Nonetheless, practitioners seem generally positive
to studies performed by the SE research community.

In 2017,
Ivanov \etal~\cite{IRSY17} investigated the gaps between research and practice
by surveying software engineers and comparing their answers to research topics
covered by recent ICSE and FSE publications.
Inconsistencies were detected between practitioners' needs and actions:
while development productivity was deemed more important than software quality,
the majority of the examined publications involve software verification and validation.
In addition,
practitioners struggle to find improved effort estimation methods.

To facilitate continuous and collaborative technology transfer,
Mikkonen \etal~\cite{MLMO18} proposed a model for large consortia
of companies and research institutes working on a common research topic.
The model was applied in two national Finnish software research programs.
To evaluate and refine it
the authors conducted interviews with four participating companies.
According to the findings,
companies perform substantial SE research
to create new business opportunities, and
appear willing to provide data to academia
for performing empirical SE research
when they identify a company benefit.
To improve industry-academia collaboration,
technology transfer models should foster co-creation and co-learning
rather than a linear one-way product transfer from academia to industry.
Research institutes can improve their mindset by
grasping long-- and short-term company needs as well as
emerging SE research trends, solutions, and results.

To address the relatively limited joint industry-academia collaborations
in SE,
Garousi \etal~\cite{GPO16} provided a set of challenges and best practices
for planning and conducting collaborative projects.
The authors denote as challenges:
difficulty in understanding the industry problems;
differences in objectives, reward systems, and useful attributes; and
difficulty in managing intellectual property rights.
Among best practices
we discern: organizing regular workshops and seminars with the industry;
assuring continuous learning from industry and academia sides;
ensuring management engagement;
grounding research on real-world problems;
demonstrating explicit benefits to the industry partners; and
maintaining agility during collaboration.

In a more recent work,
Garousi \etal~\cite{GPFF19} explore and characterize the state
of industry and academia collaborations in SE
through an opinion survey among researchers and practitioners.
Around a hundred collaborative projects from 21 countries were analyzed,
revealing that the most frequent topics are
testing, quality, process, and project management.
The vast majority of collaborative projects result in more than one publication,
while more than half have a positive impact on the industry parties,
usually through a new approach, method, technique, or tool.
To improve industry-academia collaborations,
the authors recommend a set of good practices, including
performing pilot tests in laboratory settings before industrial releases,
cultivating trustful relationships with practitioners,
investing in regular meetings to promote the team spirit, and, again,
adopting iterative approaches such as agile methods.

\subsection{Ranking Studies}
To examine the health of SE conferences,
Vasilescu \etal~\cite{VSMB14} used a metrics suite to measure the stability,
openness, representativeness, availability, and scientific prestige
of eleven conferences in a ten-year window, between 1993--2004.
Although SE conferences are generally healthy
and display high author turnover,
there are considerable differences between wide-- and narrow-scoped conferences
with regard to the aforementioned measures.
For instance,
narrow-scoped SE conferences tend to be more introvert than wide-scoped,
while maintaining more representative\footnote{Representativeness of a program committee is measured through the ratio of its members that have never co-authored a paper in the conference's past editions~\cite{VSMB14}.} program committees and lower author turnover.

In a 13-part study series spanning the years 1993--2008,
Glass \etal~\cite{Gla94,WTGB11} assessed scholars and institutions
based on the number of their publications in systems and SE.
In general, top-ranked academic institutions outnumber industrial research centers.
Although the USA was first in number of top-ranked institutions up to 2002,
it has been surpassed by the Asia-Pacific institutions since 2003.

Similarly,
Ren and Taylor~\cite{RT07} ranked individuals and organizations
according to their publications in the ICSE, FSE, TSE, and TOSEM venues between 2000--2004.
The majority of top-ranked scholars and institutions come from the USA,
while a significant number originates from Europe.
The authors argue that ``rankings based on publications can supply useful data
in a comprehensive assessment process'',
but Parnas~\cite{Par07} warns that ``measuring productivity
by counting the number of published papers slows scientific progress''.
Some organizations may be more conservative with publications.
Consider, for example,
Apple's limited publishing activity compared to its innovation~\cite{Ec22}.

A more recent assessment of top-cited SE researchers was conducted
by Petersen and Ali~\cite{PA21}
by analyzing a multi-field dataset of author citations provided
by Ioannidis \etal~\cite{IBB20}.
The authors report that
37\% of top researchers of the dataset were mistakenly assorted in the SE field,
while Barry Boehm is the leading SE author.
The majority of top SE researchers come from the USA, Canada, or the UK,
and are affiliated with Microsoft.
Along with SE,
researchers are frequently involved in artificial intelligence, image processing,
and human factors.

\subsection{Studies on Patent Metadata}
\label{sec:related-patents}
Shortly after 2000,
Agrawal and Henderson~\cite{AH02} evaluated the contribution of patents
to knowledge transfer from universities to the industry
by focusing on the MIT Departments of
Mechanical Engineering, and
Electrical Engineering and Computer Science.
Patents account for less than 10\% of knowledge transfer
from the aforementioned Departments,
while the majority of the faculty never patents at all.
Although patent volume does not predict publication volume,
it is positively correlated with paper citations,
offering insight into the impact of university research.
In general,
patenting is a complementary rather than substitutional activity
for fundamental research.

Through a systematic analysis of citation linkages
between US patents and research papers,
Narin \etal~\cite{NHO97} assessed the contribution of public science
to industrial technology.
To collect all papers cited by patents,
the authors employed a similar approach to ours
(see Section~\ref{sec:patents-citations})
by extracting all non-patent references
from around 400\,000 US patents.
In total,
73\% of papers cited by patents
originate from academic, governmental, and other public institutions,
while only 27\% were authored by industrial scientists.
Patent-to-science linkage has a strong national component
with US patents heavily citing US papers,
while the linkage is subject-specific
(\eg chemical patents cite chemistry papers).
Furthermore,
patent-cited engineering and technology papers are mainly published
in electrical engineering journals,
and IEEE is the top publisher.

\section{Methods}
\label{sec:methods}

\begin{figure*}[t]
\centering
\includegraphics[width=\textwidth]{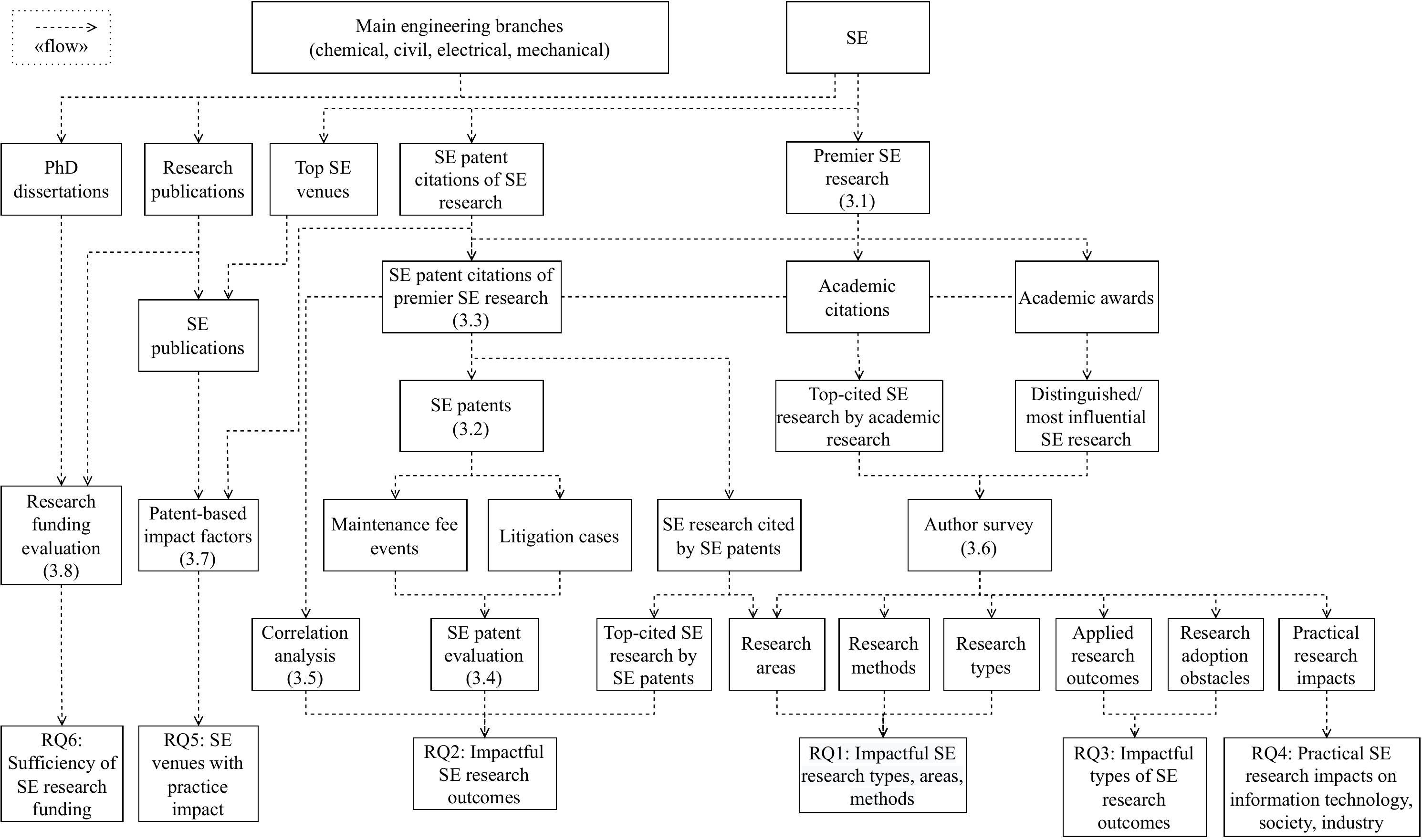}
\caption{Information flow of study methods.}
\label{fig:methods}
\end{figure*}

We framed our investigation on the impact of SE research in practice
in terms of the following research questions.

\setlength{\parindent}{0cm}
\par{\textbf{RQ1}}
\emph{What types, areas, and methods of SE research are impactful?}
To answer this,
we first collected a set of SE research papers published in leading venues
and complemented them with their assigned topics,
academic and SE patent citation counts, and awards.
Through a survey on authors of most-cited and awarded publications,
we identified impactful SE research types, areas, and methods.
The set of impactful areas was enriched
by extracting the topics of the papers cited by SE patents.
\par{\textbf{RQ2}}
\emph{What are the outcomes of impactful SE research?}
For this we retrieved the most-cited SE research by SE patents, and
evaluated the citing patents
based on their associated litigation cases and maintenance fee events.
Furthermore,
we examined the correlation between patent and academic citations,
and patent citations and academic awards.
\par{\textbf{RQ3}}
\emph{What types of SE research outcomes are impactful?}
Through a survey on authors of top-notch SE publications
we identified the practical footprint of SE research
and how its results have been exploited by the industry.
In addition,
we detected potential obstacles in the practical adoption of SE research.
\par{\textbf{RQ4}}
\emph{What are the main practical impacts of SE research
on information technology, society, and industry?}
Through the aforementioned survey
we further investigated how premier SE research changed the state of practice.
\par{\textbf{RQ5}}
\emph{Which SE venues are more likely to publish papers that impact practice?}
We assessed the practical impact associated with top SE venues
through their patent-based impact factors,
which we obtained by dividing their publication counts with their patent citations.
\par{\textbf{RQ6}}
\emph{Is SE research funding sufficient for obtaining results
that are relevant in practice?}
We approximated SE research expenditures (and thus funding)
by examining the total number of
existing SE publications and PhD dissertations, and
compared them to those of the main engineering branches to evaluate sufficiency.
\setlength{\parindent}{15pt}

\par{\textbf{Methods' Overview}}
An overview of the methods we employed to answer these six research questions is presented
through a UML information flow diagram in Fig.~\ref{fig:methods}.
Their extended descriptions are introduced in the subsequent Sections.
From the collected premier SE research (Section~\ref{sec:se-research})
we retrieved the academic citations and awards,
and SE patent citations (Section~\ref{sec:patents-citations}).
The authors of top-cited and awarded SE publications formed the sample
of our survey (Section~\ref{sec:survey}).
The set of SE patents citing the premier SE research
(Section~\ref{sec:cpc}) was evaluated
in terms of litigation and maintenance fee events (Section~\ref{sec:patents-evaluation}).
A correlation analysis was also conducted using the academic citations and awards,
and SE patent citations (Section~\ref{sec:correlation}).
From the patent-cited SE papers we extracted their research areas and top-cited papers.
Through the correlation analysis, SE patent evaluation, and top-cited SE research
by SE patents we answered RQ2.

Furthermore,
the research areas of the patent-cited SE papers, and
the research areas, methods, types,
applied research outcomes, practical impacts, and adoption obstacles
of the surveyed SE papers
provided us insights for RQ1, RQ3, and RQ4.
For RQ5
we identified the top SE venues,
retrieved their publication and SE patent citation counts,
and divided them to compute the venues' patent-based impact factors
(Section~\ref{sec:venues}).
Finally, for RQ6
we extracted the publication and PhD dissertation counts of SE
and the main engineering branches
(\ie chemical, civil, electrical, and mechanical~\cite{Ham00}),
and compared them to assess the sufficiency of provided SE research funding
(Section~\ref{sec:funding}).

\subsection{Premier SE Research}
\label{sec:se-research}
We created a dataset of research papers published
in four top-notch SE venues, namely
the \emph{International Conference on Software Engineering (ICSE)},
the \emph{IEEE Transactions on Software Engineering (TSE)},
the \emph{ACM Transactions on Software Engineering and Methodology (TOSEM)}, and
the \emph{Empirical Software Engineering (EMSE)} journal.
From these we retrieved 11\,419 papers
by downloading the complete DBLP computer science biography
database (version April 4, 2017),\footnote{\url{https://dblp.org/}}
and filtering its XML records to retain those
whose \emph{inproceedings key} tag contained \emph{conf/icse}, and those
whose \emph{article key} tag contained \emph{journals/tse}, \emph{journals/tosem},
or \emph{journals/ese}.
We excluded records with undefined \emph{title} or \emph{author} tags,
and obtained
6\,950 ICSE,
3\,417 TSE,
428 TOSEM, and
624 EMSE publications
spanning the years 1975--2017.
For each selected study, we listed
its publication year,
first author,
title, and
digital object identifier (DOI).

We then obtained the topics of the selected studies as follows.
For the TSE and the IEEE-published ICSE papers,
we used the DBLP-extracted DOIs
to download the corresponding IEEE Xplore HTML pages.
From these we extracted the INSPEC controlled terms of each paper.
In total,
we retrieved 536 distinct topics of 1\,106 TSE papers, and
547 distinct topics of 1\,348 IEEE-published ICSE papers.
(The overall distinct topics were 730.)

For the TOSEM and the ACM-published ICSE papers,
we obtained the \emph{ACM DL Abstracts and Titles for Research Purposes} database
covering the period until 2017
from the ACM Publications Operations Manager
(C. Rodkin, personal communication, October 21, 2020).
We filtered its XML metadata files to retain those
residing in directories starting with
\emph{TRANS-TSEM} or \emph{PROC-ICSE}.
We normalized the ICSE analysis to include only the regular
main track, excluding Companion and Future proceedings
as well as those of the following often co-published tracks:
New Ideas and Emerging Results;
Software Engineering Education and Training;
Software Engineering in Practice; and
Software Engineering in Society.
We removed these tracks to preserve homogeneity of our ICSE dataset,
which would otherwise be affected by the tracks' varying start years
and substantial missing data in the ACM database.
For each paper,
we extracted its assigned
2012 ACM Computing Classification System (CCS)~\cite{Rou12} concepts
from the \emph{concept\_id} and \emph{concept\_desc} tags.
Overall,
we obtained 287 distinct topics of 429 TOSEM papers, and
439 distinct topics of 1\,648 ACM-published ICSE papers.

For the EMSE papers,
we did not use Springer's keywords because they do not have a defined structure.
Instead, we retrieved their assigned 2012 ACM CCS concepts.
EMSE is not included in the aforementioned ACM database,
thus we web-scrapped the CCS concepts of the papers
from their associated ACM Digital Library web pages
by developing a Python script using the Beautiful Soup package~\cite{ZHP15}.
In this way, we collected 216 distinct topics
of 239 EMSE papers.
(The overall distinct topics for the TOSEM, ACM-published ICSE, and EMSE papers were 545.)

We also retrieved the academic citations (\ie references by other research papers)
and conference awards of the SE research papers.
Since the DBLP data did not contain citation counts,
we obtained them
from Elsevier's Scopus database\footnote{\url{https://scopus.com/}}
by querying the field \emph{Source title}
using as input each of the four venue names combined with the publisher
(\eg \emph{ACM Transactions on Software Engineering and Methodology}).
In this way,
we retrieved 13\,862 ICSE, 421 TOSEM, 3\,464 TSE, and 694 EMSE records,
again spanning the years 1975--2017.
For the awards,
we manually searched in the ICSE proceedings
for all \emph{distinguished} and \emph{most influential} papers up to 2016,
identifying a set of 74 and 29 papers, correspondingly.

\subsection{SE Patents}
\label{sec:cpc}
To identify SE-related patents,
we adopted one of the two approaches recommended by Griliches~\cite{Gri90}:
employing a patent classification system developed by a patent office.
The alternative involves reading
and manually classifying individual patents
(\eg the work by Bessen and Hunt~\cite{BH07})---this
would restrict the research scope
due to our small number of human raters.
Instead,
we used the Cooperative Patent Classification (CPC) system (version 2020.08),\footnote{\url{https://www.cooperativepatentclassification.org/}}
which has been jointly developed by the European Patent Office (EPO)
and the US Patent and Trademark Office (USPTO)~\cite{Bla12}, and
``is a further step towards a more general harmonization
of the world's patent classification systems''~\cite{EPO13}.

The CPC system is divided into nine sections,
which in turn are subdivided into
classes,
subclasses,
groups, and
subgroups---we manually looked for SE categories
in all levels of the hierarchy.
To ensure consistency of this manual process,
guidelines recommended in the work by Brereton \etal~\cite{BKBT07} were followed:
the first author of this paper performed the lookup, and
the last author validated the identified SE categories.
Specifically,
the first author identified as relevant all subgroups
under the group G06F8/00 \emph{Arrangements for software engineering},
along with any subgroups
mentioned in them
that belonged to other groups.
For instance,
G06F8/451 \emph{Code distribution} references
G06F9/5083 \emph{load rebalancing},
and G06F9/5083 contains G06F9/5088 \emph{involving task migration}---the latter
was also included in this case.
In the end,
173 SE-related categories were identified
by the first author.

The last author verified these categories
taking into account the lower-level contents
of the ACM CCS \emph{Software and its engineering} concept.
Categories associated with the following irrelevant CCS concepts were removed:
\emph{Hardware};
\emph{Distributed computing methodologies};
\emph{Concurrent computing methodologies};
\emph{Security and privacy};
\emph{Operations research}.
Furthermore, the entire CPC subgroup
G06Q10/06 \emph{administrative, planning or organization aspects of software project management}
mentioned by G06F8/00 was excluded,
because although it seems relevant to SE,
it effectively applies to any management context.
The resulting set of SE-related CPC categories has 117 members.

To retrieve patents belonging to the 117 CPC categories,
we queried the Google Patents Public Data (GPPD) dataset~\cite{GPPD17} on BigQuery.
GPPD is a worldwide bibliographic and US full-text dataset of patent publications
provided by IFI CLAIMS Patent Services and Google,
and updated on a quarterly basis---for this study,
the April 2020 version was used.
From Table \emph{patents.publications\_202004} we extracted 304\,368 distinct patents
associated with at least one of the aforementioned CPC categories.

\subsection{SE Research Cited by SE Patents}
\label{sec:patents-citations}
Scientific literature cited by patents (\ie science linkages) can provide insights
into the impact of science on industry~\cite{OECD09}.
Science linkages are usually considered the state of the art
and help in evaluating an invention's novelty and patentability.
Companies whose patents contain many science linkages are regarded closer to science,
basing their technology on scientific progress~\cite{OECD09}.
Moreover,
science linkages may be used as predictors of a company's financial performance:
high-tech companies typically surpass their competitors in science linkages~\cite{Nag07}.

Motivated by this,
we assessed the practical impact of the SE papers in terms of
their citations by SE patents.
First,
we collected all non-patent literature cited by the patents,
which is stored in plain text format
in Table \emph{patents.publications\_202004} (field \emph{citation.npl\_text}).
Querying this Table we extracted 830\,379 text references
associated with 92\,772 distinct patents.
To identify any SE papers in the references,
we followed two approaches:
DOI crosschecking, and
title and author mapping.
For the first,
we looked up the available 6\,017 ICSE, 3\,405 TSE, 426 TOSEM, and 509 EMSE DOIs
in the references,
and found
43 ICSE, 52 TSE, 12 TOSEM, and 3 EMSE papers cited
by 121 patents.
To include in the process papers and references without available DOIs,
we also searched in the references all SE titles,
combined with the last names of the first authors.
In case both a title and a name were found in a reference,
we considered this a match.
From this process,
895 ICSE, 630 TSE, 104 TOSEM, and 20 EMSE papers
(that were not identified through DOI crosschecking)
were found cited by 4\,248 patents.
In total, 1\,690
(912 ICSE, 649 TSE, 107 TOSEM, 22 EMSE) distinct papers are referenced
in 4\,354 distinct patents.

To evaluate our method,
we estimated the accuracy using a random sample of the collected references.
The sample size of a total of 6\,469 references was calculated at around 363
using Cochran's sample size and correction formula for the proportion~\cite{Coc77}
(95\% confidence, 5\% precision).
We manually verified the sample references and marked 19 (5\%) of them as false positives.
These fall into three categories:
different paper version (mostly earlier)\footnote{In patents,
the established practice is to cite the earliest version
of equally important documents~\cite{OECD09}.} cited by patent (63\%);
inherent dataset issue
(\ie wrong paper title or DOI documented in dataset---32\%);
method insufficiency
(\ie the searched DOI is a subset of the one referenced in patent---5\%).

\subsection{Evaluation of Citing SE Patents}
\label{sec:patents-evaluation}
As explained in Section~\ref{sec:methods},
we estimated the value of the SE patents that cite SE research
based on two indicators:
patent maintenance fee events and litigation cases.
These are recommended measures of patent quality
due to the associated substantial monetary expenses~\cite{Bes08}.

We extracted the litigation cases associated with the citing patents
from the USPTO Patent Litigation Docket Reports Data,
which contain detailed patent litigation information
on 81\,350 unique district court cases filed during the period 1963--2016~\cite{MTT17,SSM19}.
Specifically,
we joined the files \emph{cases} and \emph{patents} (version 2016),\footnote{\url{https://bulkdata.uspto.gov/data/patent/litigation/2016/}}
modifying patent publication numbers to follow the convention used in the GPPD dataset,
and retrieved the number of cases
as well as the corresponding aggregated demanded monetary damages per patent.
For each SE paper cited by patents (Section~\ref{sec:patents-citations}),
we listed the total number and damages of the corresponding litigation cases.

Maintenance fee events of patents granted since 1981 are provided
on a weekly basis by USPTO;\footnote{\url{https://bulkdata.uspto.gov/data/patent/maintenancefee/}}
we used the February 8, 2021 release which comprises 18\,523\,706 unique events,
again adapting patent publication numbers to the GPPD standard.
For each event, a fee code is listed, but not the actual monetary value;
we manually extracted the related fee values
from the USPTO Fee Schedule\footnote{\url{https://www.uspto.gov/learning-and-resources/fees-and-payment/uspto-fee-schedule}}
(effective since January 2, 2021) as follows.
A set of 157 codes are reported in the documentation file of the dataset.
From these we excluded
37 codes that are irrelevant to payments or refunds, and
four codes subject to the 37 Code of Federal Regulations, Paragraph 1.28,
concerning debts occurring from errors in the small entity status---these are not fixed values.
The remaining codes were mapped to their values based on the USPTO Fee Schedule,
while deprecated codes were first associated with the replacing ones
through the Fee Schedule Crosswalk, FY2002--2003.\footnote{\url{https://www.uspto.gov/learning-and-resources/fees-and-payment/fee-schedule/fee-schedule-crosswalk-fy2002-2003}}
Finally,
a total of 112 codes regarding payments of maintenance fees, surcharges, and refunds
were mapped to the current fee rates.

Similar to litigations,
for each SE paper cited by patents,
we extracted the number and aggregated monetary value
of the corresponding patent maintenance fee events.
To do this,
we computed the monetary value of each distinct patent
by summing its fee payments and surcharges, and subtracting any refunds.
Refund cases were also excluded from the total number of maintenance fee events
of each patent.

\subsection{Correlation Analysis}
\label{sec:correlation}
We investigated whether patent citations are correlated
with academic citations and academic awards.
For the 1\,690 papers cited by patents
we joined their patent citations with the academic ones,
and also marked the awarded papers.
To select the appropriate correlation coefficient,
we tested the two citation distributions for normality
with D'Agostino and Pearson's omnibus test of normality~\cite{AP73},
and found that they do not follow a normal distribution.
Therefore,
we used Spearman's rank correlation coefficient ($\rho$)~\cite{Spe1904},
which summarizes the monotonic relationship between two variables
that do not follow a normal distribution.
Patent citations constituted the dependent variable.

\subsection{Survey on SE Research Authors}
\label{sec:survey}
We conducted a survey on authors
of exceptional SE publications
following the set of ten activities recommended in
Kitchenham and Pfleeger's
six-part series of survey research principles~\cite{PK01,KP03}.

\par{\textbf{Survey Design}}
We adopted a cross-sectional, case-control, observational study design,
which means that
candidates were surveyed about their past experiences
at a fixed point in time~\cite{KP02a}.
The goal of the survey was \emph{to examine how landmark research
has affected SE practice},
therefore we framed the objectives of the survey in terms of RQ1, RQ3, and RQ4 introduced in Section~\ref{sec:methods}.

\par{\textbf{Survey Sample}}
The sample was composed of the first authors of the most-cited studies
published in ICSE, TSE, TOSEM, and EMSE
as well as studies that received distinguished and most influential paper awards.
We consider that first authors are the ones
who have contributed the most to the associated research~\cite{PA21},
and are therefore more familiar and knowledgeable about the investigated topics,
and also the most appropriate to provide the required feedback.
For each venue and year up to 2016,
we selected the five most-cited publications
(Section~\ref{sec:se-research}),
leading to a set of 613 studies.
We complemented this with the 103 awarded distinguished
and most influential ICSE papers.
After removing duplicate studies
(\ie ICSE publications that were subsequently extended in TSE, TOSEM, or EMSE)
keeping the latest occurrences,
our final set included 677 distinct papers,
which were associated with 566 distinct first authors.
From these, we managed to contact via e-mail 475 first authors
of 593 papers
(204 ICSE, 176 TSE, 121 TOSEM, 92 EMSE).
These constituted the survey sample.
Contact failures involved
missing or defunct e-mail addresses,
deceased or unavailable authors, and
rejected e-mail deliveries.

A total of 50 (out of 475---10\%) surveyed authors
associated with 58 (out of 593---10\%) papers were practitioners
when their work was published.
To compute this we retrieved the first author affiliations from Elsevier Scopus's API
through the \emph{pybliometrics} Python interface~\cite{RK19}
using the publication DOIs as input.
We excluded affiliations containing the (case-insensitive) string ``univ'',
which would most likely involve universities,
and manually looked in the remaining records for companies.
In this way we identified 50 authors affiliated with 36 companies,
including IBM, Nokia, Microsoft, Sun Microsystems, AT\&T, General Electric, Intel,
and Robert Bosch.

\par{\textbf{Survey Instrument}}
Participants were provided with personalized questionnaires
that mentioned at the beginning the author's name,
the examined paper,
the venue and year of publication,
and the reason it was selected (\ie top-cited, distinguished, or most influential).
(First authors of multiple papers were provided with multiple such questionnaires.)
They could also include and review additional publications of theirs
not included in the list,
which they considered to have made significant impact
on SE practice.
A total of five responses were collected from this option from four distinct participants.

The questionnaire was composed of mandatory and optional
open-ended, multiple choice, and Likert scale questions,
accompanied by neutral and free-text options.
Participants were initially requested to rate on a three-level Likert scale
the extent of practical impact of their publication,
and specify through a multiple choice question in what products or processes
their work has been incorporated.
In an open-ended question,
they were subsequently invited to expand on the practical impact,
along with ways in which their work changed the state of practice.
In case of absence of practical impact,
they were asked to comment on the reasons behind this.
Furthermore,
respondents were asked to select the research areas of their work
from a list with the first-- and second-level entries
to the \emph{Software and its engineering} concept of the ACM CCS~\cite{Rou12}.
Another question involved specifying the research types and methods
employed in the publication
from a list adapted from the work by Easterbrook \etal~\cite{ESSD08}.
Finally,
participants could list other impactful papers (of which they were not first authors),
leave their e-mail address to receive a report with the survey results,
and comment on the survey and its topic.

\par{\textbf{Survey Evaluation}}
Two pilot studies were conducted on candidates of the survey sample
to evaluate and refine the questionnaire.
The first pilot was internal and was completed
by three members of the laboratory
associated with four publications.
The second pilot was external and was distributed
to a random subset of eleven candidates
linked to 16 publications.
This trial was held from September 5th to 30th, 2017,
and we received six responses from six distinct participants
(55\% response rate in terms of authors, 37\% in terms of papers).

\par{\textbf{Survey Operation}}
The final survey ran from October 1st to November 5th, 2017,
and from May 18th to 31st, 2022.
The pilot and the first run of the final survey were hosted on the
\emph{SurveyGizmo} online survey platform,\footnote{\url{https://www.surveygizmo.com/}}
while the second run was provided as a \emph{Google form}.
Both runs were distributed to the candidate participants through an invitational mail.
E-mail addresses were manually fetched from the candidates' personal websites.
The mailing process was automated but retained personalization,
as explained before.
Candidates were informed about the average time required
for the questionnaire completion---around three minutes,
and the survey objective.
The final survey received in total
165 responses from 125 distinct authors
(26\% response rate in terms of authors, 28\% in terms of papers).
The anonymized responses are included in the provided dataset.
Some answers were redacted upon respondents' request.

\par{\textbf{Survey Analysis}}
We applied manual coding~\cite{CS90} to summarize the answers
to the four open-ended questions and
the free-text option of a multiple choice question.
For each question,
the second and third authors of this paper split the answers in two sets,
and each individually applied codes to a half (in a shared online spreadsheet).
At least one and up to six codes were applied to each answer.
Next,
the first author grouped together conceptually-related codes
by generalizing or specializing them,
following the Qualitative Content Analysis approach~\cite{LSMH15}.

\subsection{SE Venues}
\label{sec:venues}
We assessed the practical impact of top SE venues
by computing their patent-based impact factors as follows.
We retrieved the Google Scholar Metrics list of top publications
under \emph{Categories$>$Engineering \& Computer Science$>$Software Systems}
(July 2021 index),\footnote{\url{https://web.archive.org/web/20211118180610/https://scholar.google.gr/citations?view_op=top_venues&hl=en&vq=eng_softwaresystems}}
excluding the following entries that center on programming languages and algorithms:
PLDI,
POPL,
Proceedings of the ACM on Programming Languages,
TACAS,
PPOPP.
For the remaining venues we extracted their publication counts from Elsevier Scopus
for a ten-year window between 2009--2019.
This window was selected to approximate publications of all years,
avoiding bias due to different venue start dates,
while the particular range was chosen to align with the GPPD version.
Therefore,
we queried all venue names combined with the year range,
restricting document types to
reviews and articles for journals, and
papers for conferences,
similar to Clarivate's impact factor calculation~\cite{IF18}.
To retrieve the patent citation counts of the SE venues,
we searched their full names, abbreviations, and acronyms
in the 830\,379 non-patent references of patents (Section~\ref{sec:patents-citations}).
Finally,
citation counts were divided with publication counts
to calculate the patent-based impact factors.

\subsection{SE Research Funding}
\label{sec:funding}
We also approximated the sufficiency of provided SE research funding
by comparing it to that of the main engineering branches.
We initially searched for existing empirical analyses in the literature,
but did not obtain any fruitful results.
As a workaround,
we approximated funding based on the number of existing publications
and PhD dissertations,
given that both activities are usually grant-aided.
We compared SE results to those of the four main engineering branches:
civil, mechanical, electrical, and chemical~\cite{Ham00}.
For publications,
we extracted from Scopus all English papers between 2010--2020
that belong to the subject areas
\emph{Engineering},
\emph{Computer Science}, or
\emph{Chemical Engineering},
and contain the above engineering fields in the keyword list.
For dissertations,
we queried the Open Access Theses and Dissertations (OATD)
database,\footnote{\url{https://oatd.org/}}
and retrieved all English PhD dissertations from the same period,
whose subject and discipline are the corresponding engineering fields
(except for SE,
where we specified computer science as discipline).

We used OATD for the following reasons.
The database contains more than six million electronic theses and dissertations (ETDs)
published between 1973--2022 (retrieved April 20, 2022)\footnote{\url{https://web.archive.org/web/20220420032348/https://oatd.org/oatd/search?q=*\%3A*&sort=date}}
from about 1\,100 universities, colleges, and research institutes~\cite{Bai22}.
OATD is considered one
``of the finest resources to access ETDs worldwide''~\cite{Bai22},
and is a recommended international thesis resource for researchers
by the Networked Digital Library of Theses and Dissertations (NDLTD)~\cite{NDLTD}.
In a recent evaluation of four prominent search engines on retrieving ETDs
through various search techniques including title, keyword, and author search,
OATD was ranked second in overall performance (88.5\%),
closely following Google (89\%),
and surpassing Yahoo (78\%), and Google Scholar (76\%)~\cite{LPK22}.
Furthermore, OATD has been used in various research studies.
These mainly evaluate the quality of the database
and its repositories~\cite{Bai22,LPK22,Wan19},
investigate user behavior~\cite{Coa14},
assess and contrast cultural heritage~\cite{Gre14,AJ21}, and
identify the benefits and impact of open access PhD e-theses~\cite{FPVR16}.

\section{Results}
\label{sec:results}
In this section we present the study findings
from the patent and survey analysis,
in respect to the research questions described in Section~\ref{sec:methods}.
Survey percentages are calculated on the basis of responses (165)
rather than distinct authors (125).
Codes derived from the manual coding process
of the open-ended answers are set in \oecode{bold}.

\subsection{RQ1: Impactful Research Types, Methods, Areas}
\label{sec:rq1}
Among survey participants,
40\% (66 participants)
consider that the work described in their paper resulted in
\emph{some} practical impact and 29\% (47) in \emph{wide} practical impact,
as opposed to 21\% (35) who do not believe that their work had any practical impact,
followed by 10\% (17) who are unaware of any footprint.
The variety in the responses could be explained by
the entailed subjectivity,
the fact that awards and citations are, by definition, not exclusively tied
to practical impact, or
the authors' different impact expectations.
For instance,
authors may be more ambitious than award-givers and researchers citing their work.

Table~\ref{tab:types} displays the practical impact
of SE research types
declared in the collected responses.
Types are sorted in descending order of appearance frequency.
Empirical research (\eg investigating the adoption of engineering methods,
developing new tools)
appears the most common type among the examined publications,
followed by design research (\eg developing new methods)
and theoretical research (\eg proving properties of systems axiomatically).
Papers that were characterized by some or wide impact were grouped together
as \emph{impactful}.
All types proved considerably impactful.

Table~\ref{tab:methods} presents the practical impact of the sample's employed research methods
in descending frequency order.
Action research (\eg being embedded in the development team) and
exploratory research (\eg describing characteristics of a population or phenomenon
under investigation) are
the most impactful ones.
Case study (\eg applying a new technique on existing systems) is the most frequent method
but less impactful.
In addition to our survey's predefined set,
the following methods were also reported for the examined publications:
\oecode{survey},
\oecode{secondary research}
(\eg systematic review, literature review, meta-analysis),
\oecode{replication study},
\oecode{content analysis},
\oecode{econometric analysis},
\oecode{data collection and analysis},
\oecode{formal theory}, and
\oecode{design and evaluation}.
Both research types and methods can overlap,
because papers may employ many of them.

\begin{table}[!t]
\renewcommand{\arraystretch}{1.3}
\caption{Practical Impact of SE Research Types\\(according to survey respondents' self-evaluation)}
\label{tab:types}
\centering
\resizebox{0.3\textwidth}{!}{
\begin{tabular}{l r r r}
\hline
\textbf{Type}
            & \textbf{Papers}
                    & \textbf{Impactful}
                            & \textbf{\%}\\
\hline
Empirical   & 119   & 85    & 71\\
Design      & 72    & 53    & 74\\
Theoretical & 34    & 26    & 76\\
\hline
\end{tabular}}
\end{table}

\begin{table}[!t]
\renewcommand{\arraystretch}{1.4}
\caption{Practical Impact of SE Research Methods}
\label{tab:methods}
\centering
\resizebox{0.44\textwidth}{!}{
\begin{tabular}{l r r r}
\hline
\textbf{Method}
                                            & \textbf{Papers}
                                                    & \textbf{Impactful}
                                                            & \textbf{\%}\\
\hline
Case study                                  & 85    & 63    & 74\\
Controlled/Natural experiment               & 48    & 32    & 67\\
Exploratory research                        & 31    & 25    & 81\\
Action research                             & 18    & 16    & 89\\
Ethnography                                 & 15    & 10    & 67\\
Simulation                                  & 10    & 8     & 80\\
Other                                       & 39    & 21    & 54\\
\hline
\end{tabular}}
\end{table}

\begin{table}[!t]
\renewcommand{\arraystretch}{1.4}
\caption{Practical Impact of SE Research Areas}
\label{tab:areas}
\centering
\resizebox{0.48\textwidth}{!}{
\begin{tabular}{l r r r}
\hline
\textbf{ACM CCS SE Area~\cite{Rou12}}
                                       & \textbf{Papers}
                                               & \textbf{Impactful}
                                                       & \textbf{\%}\\
\hline
Software creation and management       & 142   & 104   & 73\\
Software organization and properties   & 106   & 77    & 73\\
Software notations and tools           & 93    & 71    & 76\\
\hline
\end{tabular}}
\end{table}

\begin{table*}[!t]
\renewcommand{\arraystretch}{1.4}
\caption{Practical Impact of SE Research Subareas}
\label{tab:subareas}
\centering
\resizebox{\textwidth}{!}{
\begin{tabular}{l l r r r}
\hline
\textbf{ACM CCS SE Area~\cite{Rou12}}
    & \textbf{Subarea}
        & \textbf{Papers}
                & \textbf{Impactful}
                        & \textbf{\%}\\
\hline
Software creation and management
    & Software verification and validation
        & 62    & 47    & 76\\
    & Software development techniques
        & 58    & 44    & 76\\
    & Designing software
        & 48    & 37    & 77\\
    & Software development process management
        & 38    & 29    & 76\\
    & Software post-development issues
        & 32    & 26    & 81\\
    & Collaboration in software development
        & 20    & 14    & 70\\
    \vspace{0.1cm}
    & Search-based software engineering
        & 11    & 9     & 82\\
Software organization and properties
    & Extra-functional properties
        & 61    & 45    & 74\\
    & Software functional properties
        & 49    & 38    & 78\\
    & Software system structures
        & 38    & 26    & 68\\
    \vspace{0.1cm}
    & Contextual software domains
        & 6     & 5     & 83\\
Software notations and tools
    & Software maintenance tools
        & 44    & 36    & 82\\
    & Development frameworks and environments
        & 38    & 36    & 95\\
    & System description languages
        & 29    & 23    & 79\\
    & Software configuration management and version control systems
        & 20    & 15    & 75\\
    & General programming languages
        & 19    & 12    & 63\\
    & Formal language definitions
        & 14    & 9     & 64\\
    & Software libraries and repositories
        & 14    & 10    & 71\\
    & Context specific languages
        & 11    & 9     & 82\\
    & Compilers
        & 6     & 5     & 83\\
\hline
\end{tabular}}
\end{table*}

The impact of SE research areas and subareas
based on the survey findings can be deduced
from Tables~\ref{tab:areas} and~\ref{tab:subareas}, correspondingly.
Areas and subareas correspond to the first-- and second-level entries
to the \emph{Software and its engineering} concept
of the ACM CCS,
respectively.
From Table~\ref{tab:areas},
the overall most impactful area is \emph{Software notations and tools},
but the most frequent one is \emph{Software creation and management}.
The most impactful subarea (Table~\ref{tab:subareas}) is
\emph{Development frameworks and environments},
followed by
\emph{Contextual software domains} (\eg operating systems) and
\emph{Compilers}
with the fewest publications.
The most frequent subareas are
\emph{Software verification and validation} and
\emph{Extra-functional properties}
with moderate impact.
Again,
both areas and subareas can overlap.

We further retrieved the most frequent IEEE INSPEC controlled terms and ACM CCS concepts
of the SE papers that are cited by SE patents (Section~\ref{sec:se-research}).
We retrieved the topics of
277 ACM-- and
292 IEEE-published ICSE papers,
107 TOSEM,
227 TSE, and
14 EMSE papers.
The ten most common INSPEC terms are
\emph{program testing} (24\%---124),
\emph{software maintenance} (16\%---81),
\emph{object-oriented programming} (13\%---69),
\emph{program debugging} (13\%---68),
\emph{program diagnostics} (12\%---65),
\emph{software engineering} (11\%---57),
\emph{Java} (11\%---56),
\emph{formal specification} (10\%---54),
\emph{software tools} (9\%---47), and
\emph{software quality} (8\%---42).
Similarly,
the ten most frequent CCS concepts are
\emph{Software testing and debugging} (29\%---116),
\emph{Software development process management} (19\%---77),
\emph{Software management} (14\%---55),
\emph{Designing software} (12\%---49),
\emph{Software maintenance} (12\%---49),
\emph{Formal software verification} (11\%---43),
\emph{Program verification} (11\%---43),
\emph{Software design techniques} (\%---42),
\emph{Development frameworks and environments} (10\%---40), and
\emph{Software creation and management} (10\%---39).

\subsection{RQ2: Outcomes of Impactful SE Research}
\label{sec:rq2}
A total of 1\,690 SE papers have been cited
by 4\,354 SE patents;
the 20 most-cited ones are summarized in Table~\ref{tab:top-cited}.
The majority were published in ICSE between 1984--2013.
Through a correlation analysis between patent and academic citations,
and patent citations and academic awards (Section~\ref{sec:correlation}),
Spearman's $\rho$ was calculated at 0.25 and 0.07, respectively,
suggesting a weak positive correlation in both cases.

\begin{table*}[!t]
\renewcommand{\arraystretch}{1.4}
\caption{Most-cited SE Papers by SE Patents}
\label{tab:top-cited}
\centering
\resizebox{\textwidth}{!}{
\begin{tabular}{r l l l r r r r r r}
\hline
\#	& \textbf{Title}
		& \textbf{Authors}
			& \textbf{Venue}
					& \textbf{Year}
							& \textbf{\thead{SE Patent\\Citations}}
									& \textbf{\thead{Patent\\Lit.\\Cases}}
											& \textbf{\thead{Patent\\Maint.\\Fee\\Events}}
													& \textbf{\thead{Patent\\Maint.\\Fees (\$)}} \\
\hline
1	& \thead{Tracking down Software Bugs Using Automatic Anomaly\\Detection}
    	& Hangal and Lam~\cite{HL02}
        	& ICSE  & 2002  & 45    & -     & 34    & 64\,830 \\
2	& EDMAS: A Locally Distributed Mail System
    	& Almes \etal~\cite{ABBW84}
        	& ICSE  & 1984  & 43    & 3     & 67    & \textbf{260\,360} \\
3	& Model-based Development of Dynamically Adaptive Software
    	& Zhang and Cheng~\cite{ZC06}
        	& ICSE  & 2006  & 43    & -     & 37    & 75\,760 \\
4	& Software Deployment, Past, Present and Future
    	& Dearle~\cite{Dea07}
       		& ICSE  & 2007  & 40    & -     & 21    & 43\,760 \\
5	& Aspect-oriented Programming
    	& Kiczales~\cite{Kic05}
        	& ICSE  & 2005  & 39    & -     & 42    & 114\,290 \\
6	& Program Slicing
    	& Weiser~\cite{Wei84}
        	& TSE   & 1984  & 35    & -     & 30    & 62\,680 \\
7	& The Eden System: A Technical Review
    	& Almes \etal~\cite{ABLN85}
        	& TSE   & 1985  & 34    & 2     & 79    & \textbf{312\,580} \\
8	& Software Engineering Economics
    	& Boehm~\cite{Boe84}
        	& TSE   & 1984  & 33    & -     & 25    & 70\,340 \\
9	& Distribution and Abstract Types in Emerald
    	& Black \etal~\cite{BHJL87}
        	& TSE   & 1987  & 33    & 2     & 77    & \textbf{306\,820} \\
10	& \thead{A Cooperative Approach to Support Software Deployment\\Using the Software Dock}
    	& Hall \etal~\cite{HHW99}
        	& ICSE  & 1999  & 32    & 1     & 44    & \textbf{150\,360} \\
11	& Automated Software Test Data Generation
		& Korel~\cite{Kor90}
			& TSE	& 1990	& 32	& -		& 35	& \textbf{130\,470} \\
12	& Safe Software Updates via Multi-version Execution
		& Hosek and Cadar~\cite{HC13}
			& ICSE	& 2013	& 31	& -		& 14	& 26\,500 \\
13	& Predicting Source Code Changes by Mining Change History
		& Ying \etal~\cite{YMNC04}
			& TSE	& 2003	& 29	& -		& 14	& 31\,780 \\
14	& Call Path Profiling
		& Hall~\cite{Hal92}
			& ICSE	& 1992	& 27	& -		& 54	& \textbf{154\,180} \\
15	& \thead{Hipikat: Recommending Pertinent Software Development\\Artifacts}
		& \thead{\v{C}ubrani\'{c} and\\Murphy~\cite{CM03}}
			& ICSE	& 2003	& 27	& -		& 22	& 46\,020 \\
16	& \thead{PROVIDE: A Process Visualization and Debugging\\Environment}
		& Moher~\cite{Moh88}
			& TSE	& 1988	& 27	& -		& 38	& \textbf{134\,230} \\
17	& Refactoring
		& Fowler~\cite{Fow02}
			& ICSE	& 2002	& 27	& -		& 18	& 50\,500 \\
18	& The \emph{Pan} Language-Based Editing System
		& Ballance \etal~\cite{BGV92}
			& TOSEM	& 1992	& 27	& -		& 46	& \textbf{164\,000} \\
19	& \thead{Recovering Traceability Links in Software Artifact\\Management Systems Using Information Retrieval Methods}
		& Lucia \etal~\cite{LFOT07}
			& TOSEM	& 2007	& 26	& -		& 8	& 16\,000 \\
20	& An Intrusion-Detection Model
		& Denning~\cite{Den87}
			& TSE	& 1987	& 25	& 1		& 29	& 70\,140 \\
\hline
\end{tabular}}
\vspace{1ex}
{\raggedright \\ Papers with \textbf{bold} fees are among the top ten with the highest fees. \par}
\end{table*}

From the patents' evaluation in terms of litigations
(Section~\ref{sec:patents-evaluation}),
twelve patents citing
15 papers are associated
with 20 litigation cases
concerning patent infringements,
and five of these papers are included in the 20 most-cited ones
(Table~\ref{tab:top-cited}).
No damages are documented for the 20 cases
in the USPTO Patent Litigation Docket Reports Data,
but we manually looked them up online
and retrieved results for 14 of them:
six cases' damages are undisclosed,
while the remaining eight cases' range
from six million to eight billion dollars.
Looking at the involved parties,
the majority are/were big corporations, including
Apple, Kodak, Ericsson, Facebook, Google, Lenovo, Microsoft, Oracle, Yahoo!,
Sun Microsystems, and Radware.

Regarding renewals,
1\,953 (45\%) patents citing
1\,159 (69\%) papers are linked with maintenance fee events.
Eight papers of the top ten in fee expenses
are also among the 20 most-cited.
The number of maintenance fee events of patents citing papers
with an equivalent number of patent citations is higher for older ones,
which is reasonable as the related patents may have been renewed more times.
Although, in general,
the number of maintenance fee events seems to be associated with the aggregated fee value,
there are substantial differences in some cases.
For example, although the number of maintenance fee events is equivalent for the works
by Zhang and Cheng~\cite{ZC06}, and Moher~\cite{Moh88},
the aggregated fees of the latter are almost double that of the former.
Such discrepancies could be explained by the different entity status of the citing patents:
fees for large-entity patents are twice that of small-entity,
and quadruple that of micro-entity.\footnote{\url{https://www.uspto.gov/learning-and-resources/fees-and-payment/uspto-fee-schedule}}

\subsection{RQ3: Impactful Types of SE Research Outcomes}
\label{sec:rq3}
\begin{figure}[t]
\centering
\includegraphics[width=\linewidth]{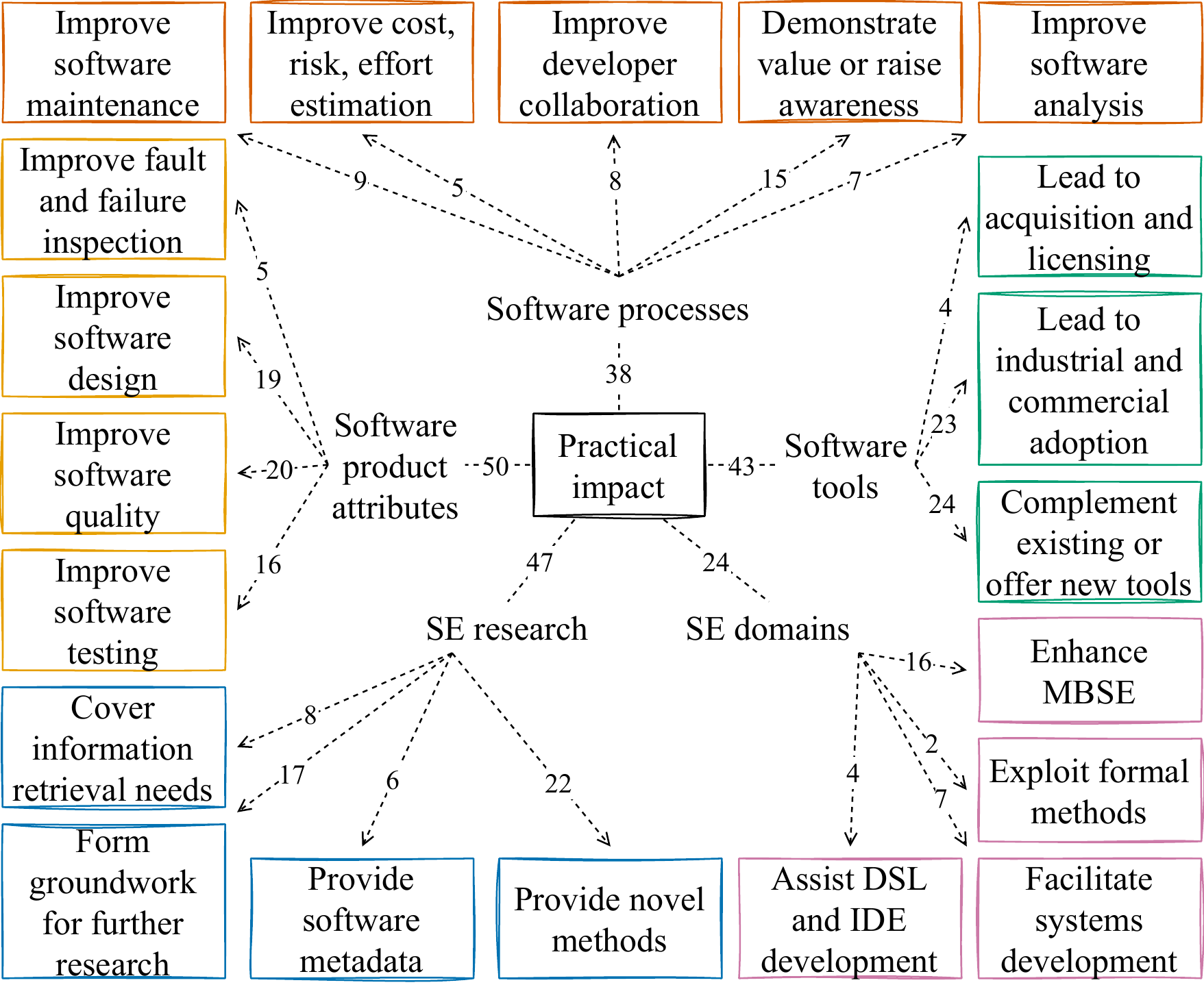}
\caption{Practical impact of SE research.}
\label{fig:impact}
\end{figure}

The results of
44\% (72) of the examined publications of the survey were incorporated
in software industry processes, practices, or methods,
41\% (68) in software development tools,
21\% (35) in marketable products,
13\% (22) in marketable services, and
19\% (31) in other areas.
In Fig.~\ref{fig:impact} we present an overview
of the practical impact of SE research grouped by type of outcome,
as expressed in the responses to the complementary open-ended question.
For each category we show the number of involved responses as weights.
An extensive review is included below,
enhanced with example quotes to support our interpretation of the responses~\cite{Kri12}.
Example quotes are marked with an \rid{R\textit{X}} notation,
where \textit{X} refers to the respondent's identification number.

From the open-ended responses
we discern the improvement
of software product attributes, particularly
\oecode{design},
\oecode{testing}, and
\oecode{quality}.
In their studies, researchers introduced methods and tools for
resolving conflicting requirements \rid{R190}, and
documenting and understanding a system's design evolution
through flexible solutions (\eg \rid{R48,R195,R400}),
supported assertion checking \rid{R178},
test approach and testing technique selection \rid{R244,R246,R320},
model-based testing and prioritization \rid{R214}, and
applied techniques such as combinatorial testing \rid{R157}
and symbolic execution \rid{R154}.
Quality was enhanced through fault injection and
clone detection tools (\eg \rid{R164,R193}),
and by analyzing software \oecode{faults and failures}.
In this way,
researchers proposed refined experiment and testing techniques (\eg \rid{R179,R199}),
proved that
combinatorial testing \response{can provide assurance effectively equivalent}
to exhaustive testing \rid{R157},
and also accelerated fault localization \rid{R100}.

Along with product attributes,
we also distinguish the advancement of software processes,
which provide support throughout a software product life cycle~\cite{swebok}.
\oecode{Maintenance} was improved by
\response{increasing the visibility of software refactoring research} \rid{R52},
\response{providing candidate patches to defects} \rid{R153}, and
by warning developers about test smell issues \rid{R317}.
\oecode{Development collaboration} was enhanced
through the advancement of code review tools \rid{R222}
and the assessment of global software team configurations \rid{R76}.
\oecode{Analysis}-wise, a study's empirical results
\response{motivated many teams to adopt static analysis} \rid{R170},
while other efforts found application in metaprogramming,
process and software analysis (\eg \rid{R1,R49,R218,R223,R251}).
To estimate \oecode{cost, risk, and effort},
researchers published effective models and metrics \rid{R133,R211}, and
enhanced analogy-based reasoning tools \rid{R173}.
Meanwhile, the objective of some studies was
to \oecode{demonstrate value or raise awareness} about a topic or existing work.
For instance,
\rid{R112} aimed \response{to alert practitioners for the need for assessing complexity},
\rid{R115} performed an empirical validation on quality measures,
which \response{are now incorporated into static analysis tools,
quality assurance practices, and as quality level agreements in software contracts},
and \rid{R136} \response{raised awareness of software development and evolution
as an economic activity}.

Significant impact was made on existing or new \oecode{tools}.
The most notable contributions concern tools for
static program analysis \rid{R49},
logic model checking \rid{R90},
software fault and defect prediction (\eg \rid{R111,R148,R301,R302,R306}),
symbolic simulation \rid{R154}, and
clone detection \rid{R164}.
Some of these tools underwent \oecode{industrial and commercial adoption}
from startups \rid{R51,R106,R156,R207} to established corporations,
such as Microsoft \rid{R222}, ABB Corporation Research \rid{R196}, BlackBerry \rid{R211},
Huawei \rid{R188}, Rational Software \rid{R224}, and IBM \rid{R251}.
\oecode{Acquisition and licensing} of tools, techniques, and even startups,
decreased the acquirers' software development costs, and
increased substantially the inventors' revenue.
As \rid{R224} declares,
they \response{eventually had almost one billion in annual revenue}.
Moreover,
the acquired work is now used globally by numerous people
(\eg \rid{R106}'s startup was acquired by Facebook, leading to an approximate monthly impact
on \response{more than two billion people worldwide}).

Some efforts directly impacted a particular domain.
To assist domain-specific language (\oecode{DSL}) \oecode{and IDE development},
researchers produced a visual language
for software modeling and specification \rid{R7},
\response{encouraged more research on preprocessors},
and experienced adoption of their work by open source frameworks
for feature-oriented software development \rid{R215}.
Exploiting \oecode{formal methods},
ad hoc pragmatic reuse tasks were simplified \rid{R228}, and
bounded model checking of multi-threaded software was improved,
amplifying the analysis of larger problems and
\response{reducing the verification time over state-of-the-art techniques} \rid{R97}.

In the field of \oecode{systems development},
we distinguish contributions to
version control \rid{R57},
concurrent \rid{R90},
hybrid \rid{R154},
service-oriented \rid{R250}, and
embedded software systems \rid{R206}.
A frequently occurring method is
model-based systems engineering (\oecode{MBSE}).
Contributions to this involve
component models \rid{R68,R242},
consolidated process models
\response{offering the ability to streamline process analysis and redesign work} \rid{R223},
and modeling language semantics of hybrid systems \rid{R154}.

Various studies affected subsequent research.
Some formed the \oecode{groundwork for further research},
obtaining multiple academic references \rid{R5,R6,R7,R197}.
As \rid{R116} states, their work on software process models
\response{blew up the foundation of clean top-down models
and forced consideration of real world issues}.
Some \oecode{novel methods} concern
qualitative research methods aiming to
\response{help researchers immerse themselves more fully in practice} \rid{R163},
goal orientation in requirements engineering \rid{R189},
experience base automation \rid{R243}, and
standardized methods (\eg drawing binary trees \rid{R70},
collecting data \rid{R234},
performing manual inspections \rid{R316}).
By providing \oecode{software metadata},
researchers managed to \response{reduce the cost of widely used software practices} \rid{R54}
and to increase analysts' efficiency \rid{R1,R227,R309}.
To cover \oecode{information retrieval needs},
researchers proposed new research agendas \rid{R166},
influenced source code search solutions \rid{R196}, and
identified substantial mismatches between IDE designs and information seeking \rid{R102,R103}.

\par{\emph{Lack of practical impact:}}
Respondents who reported no practical impact of their work
were subsequently encouraged to determine the reasons for this inefficiency.
Some efforts that could have been impactful did not succeed
due to an \oecode{immature phase},
\oecode{inefficiencies},
or additional required \oecode{effort and resources}.
In some cases,
a longer time horizon was needed to cause an impact \rid{R213,R216}.
As \rid{R213} explains,
their paper \response{is at the intersection of programming language (PL)
and SE research,
and PL research has a longer time horizon than typical SE research}.
Concerning inefficiencies, some papers contained
wrong assumptions \rid{R47},
tooling issues \rid{R64},
undocumented practical uses \rid{R149},
incomplete resolution of the addressed problem \rid{R226},
and rarely employed research methods \rid{R118}.
Furthermore,
some research outcomes have not been used by practice due to
high implementation and maintenance costs \rid{R226},
risk aversion for technology commercialization \rid{R81}, and
paucity of maintenance by their creators \rid{R64}.

Additional causes involve \oecode{undetected needs}
and lack of \oecode{support by SE practice}.
In \rid{R225}'s words,
\response{most projects feel they are doing `good enough'
with their existing processes}, and
\response{a project will not invest the time and effort
[into integrating a process such as triage recommendation]
unless there is a perceived significant benefit}.
Regarding support,
some participants expressed their concerns about SE practice
neglecting research,
for example, in requirements engineering \rid{R141} and
defect density estimation \rid{R321},
while others recognized high barriers of work adoption
in programming languages \rid{R213},
and tool integration challenges in the current SE community \rid{R144}.
It comes as no surprise, therefore,
that trending software development tools in worldwide practice
have industrial provenance~\cite{DevEco21,SO21}.
The academic and industrial research lab foundations of a few popular
version control and programming language
technologies are by now many decades old~\cite{Roc75,Tic82,RJLK78,Str83}.

Although some were \oecode{unaware} of any incorporation of their results
into development tools, products, or services \rid{R120,R304},
the majority of responses referred to an \oecode{indirect research impact}.
This includes fundamental research contributions \rid{R79,R80,R94},
for instance, to distributed systems \rid{R167},
surveys \rid{R91,R96,R304},
research methods for conducting computational experiments \rid{R75,R172},
guidelines for empirical research \rid{R161,R300,R314},
design concepts \rid{R212},
as well as contributions that \response{do not work on real systems} \rid{R185,R186,R315}.
As \rid{R216} mentions,
having no practical impact is not necessarily \response{a bad thing},
especially when a work influences other researchers,
or facilitates the exploration of different fields
and the development of techniques.

\subsection{RQ4: Practical Impacts of SE Research on Information Technology, Society, and Industry}
\label{sec:rq4}
The authors' responses to the related open-ended question
revealed various changes of SE research in the state of practice.
Through their \oecode{prototypes},
practitioners advanced \oecode{production technology}
and facilitated the \oecode{open source community}.
Specifically, they improved practices involving
mining design patterns from source code \rid{R48},
code reviewing \rid{R222},
clone detection \rid{R169},
reverse engineering and program understanding \rid{R171},
systems development \rid{R90}, quality and robustness \rid{R175,R193},
testing and analysis \rid{R157,R209,R210},
model checking \rid{R159},
remodularization optimization \rid{R62}, and
analogy-based reasoning \rid{R173,R174}.
Various prototypes were quickly adopted by open source projects \rid{R164,R195}.
Software development companies incorporated researchers' approaches
into their processes \rid{R173,R174,R209} and product lines (\eg \rid{R210}).
We also observe adoption by competitors.
Although this is \response{not exactly desirable from a company perspective,
at least the research had an impact} \rid{R170}.
Collaborations with companies benefited both parties, regardless;
companies improved their solutions,
while individuals promoted their open source tools
gaining numerous downloads \rid{R173,R174}.

Several attempts were made towards extending the \oecode{functionality}
of existing open source and proprietary products.
Researchers provided
solutions to code recommendation projects \rid{R229},
algorithms to graph layout tools \rid{R303},
support for free text search to code search engines \rid{R166,R196},
conflict modeling techniques to commercial tool sets \rid{R190},
bidirectional streaming to HTTP/2 and relevant frameworks \rid{R207},
multi-threaded software verification methods to state-of-the-art tools \rid{R97}, and
online tools for visualizing map representations of GitHub code clones \rid{R164}.
Version control systems were incorporated into Unix system distributions,
eliciting their widespread adoption by universities
and technology-leading corporations---\response{the trend setters} \rid{R57}.
Moreover,
the identification of performance bug patterns
and the development of static performance checkers allowed Android developers
to \response{generate real-time warnings when they are writing code} \rid{R127},
while the construction of rules matrices helped companies
\response{to differentiate under which legal conditions a dataset can be used
for analysis} \rid{R156}.

Along with functionality,
practitioners improved products' \oecode{design and modeling},
\oecode{quality}, and \oecode{performance}.
We observe new architecture concepts \rid{R206},
software modeling languages \rid{R152},
hybrid system modeling plug-ins \rid{R154}, and
project management frameworks \rid{R76}.
Support was provided for identifying countermeasure requirements
by modeling and analyzing threats at the application level \rid{R187},
and for encompassing visual formalisms, including statecharts,
in modeling processes \rid{R220}.
Quality-wise, some products upgraded
the robustness of cloud management platforms \rid{R193},
the measurement and decrease of attack surfaces in software systems \rid{R162},
the long-term evaluation of regression testing techniques \rid{R155},
and the independent verification and validation of proprietary tools \rid{R227}.
Performance-wise,
researchers accelerated testing \rid{R210} and
root cause analysis \rid{R251}, and
suggested solutions for extending the lifetime of products
while reducing their maintenance costs \rid{R197}.

Some studies aimed to
advance \oecode{development processes},
release \oecode{new methods}, or
educate researchers and improve their \oecode{professional practices}.
For instance,
factor-covering array generation tools impacted
statistical testing and analysis \rid{R157},
model checking techniques affected code analysis \rid{R159},
object-oriented design metrics improved software development feedback \rid{R239},
data analysis was enhanced by linking code reviews to commits \rid{R222},
analogy-based reasoning improved prediction and estimation \rid{R173}, and
goal-oriented risk analysis impacted requirements engineering \rid{R188,R189,R190}.
In short, new methods entail
approaches for developing and refining software development processes \rid{R116,R246,R250,R301,R302,R306}
and domain-specific languages \rid{R218},
model-based requirements \rid{R184},
model checking \rid{R159},
formalism-oriented abstraction levels \rid{R138},
conflict modeling \rid{R190},
syntactic preprocessors for implementing variability \rid{R215},
configuration management for component models \rid{R152},
fault injection \rid{R193}, and
accident models for safer systems \rid{R6}.
Proposed practices involve
abandoning \response{information retrieval-based traceability link recovery approaches
due to low performance} \rid{R69} and copy-pasting \rid{R51},
favoring \response{minimization of coupling} \rid{R126},
carefully testing code segments that are predicted to contain faults \rid{R111},
using software metrics to predict programmers' performance \rid{R115},
being careful with attack surfaces of software systems \rid{R162},
preferring alternative metrics to test coverage \rid{R221}, and
detecting and mitigating \response{architectural mismatch} \rid{R99}.

Although some studies may not have directly changed the state of practice
through prototypes and product add-ons,
they still influenced \oecode{follow-up work} and had a \oecode{research impact}.
Second-order mutation algorithms affected mutation testing tools \rid{R142},
a work \response{served as a guideline for reusing components} \rid{R242},
various doctoral dissertations were inspired by research
on reverse engineering and program comprehension \rid{R171},
invalid techniques deployed in safety-critical systems were revealed \rid{R183}, and
the effect of design decisions on inspection performance was evaluated \rid{R316}.
Experiments were used as a baseline for subsequent research
in aspect-oriented programming \rid{R124},
specification models led to further academic research and funding \rid{R214},
and some works were \response{instrumental in ushering in the current era
of data-driven thinking in SE} \rid{R54}.
Finally, some publications led to author \oecode{promotion} \rid{R116}.

\subsection{RQ5: SE Venues with Practice Impact}
\label{sec:rq5}
The patent-based impact factors of the top SE venues are presented
in Table~\ref{tab:venues} in descending order,
complemented with the corresponding number of publications and SE patent citations.
Journals are denoted with \emph{J} and conferences with \emph{C}.
The most impactful venue appears to be the IEEE TSE,
followed by the International Conference on Mining Software Repositories (MSR) and
the ACM SIGSOFT International Symposium on Software Testing and Analysis (ISSTA).

\begin{table}[t]
\renewcommand{\arraystretch}{1.4}
\caption{Practical Impact of SE Venues}
\label{tab:venues}
\centering
\resizebox{0.48\textwidth}{!}{
\begin{tabular}{c l l r r r}
\hline
\textbf{Type}
	& \textbf{Name}
					& \textbf{Publisher}
							& \textbf{Papers}
									& \textbf{\thead{SE Patent\\Citations}}
											& \textbf{\thead{Impact\\Factor}}\\
\hline
J	& IEEE Trans. Softw. Eng.	& IEEE		& 681		& 1\,988	& 2.919 \\
C	& MSR				& IEEE/ACM	& 469		& 556		& 1.185 \\
C	& ISSTA				& ACM SIGSOFT	& 435		& 473		& 1.087 \\
J	& Softw. Pract. Exp.		& Wiley		& 666		& 709		& 1.064 \\
C	& ICSE				& ACM/IEEE	& 3\,367	& 2\,497	& 0.742 \\
J	& IEEE Software			& IEEE		& 1\,136	& 833		& 0.733 \\
C	& FSE				& ACM SIGSOFT	& 1\,061	& 614		& 0.579 \\
C	& ICSME				& IEEE		& 940		& 511		& 0.544 \\
C	& ASE				& IEEE/ACM	& 1\,120	& 596		& 0.532 \\
C	& SANER				& IEEE		& 746		& 355		& 0.476 \\
J	& J. Syst. Softw.		& Elsevier	& 1\,979	& 364		& 0.184 \\
J	& Inf. Softw. Technol.		& Elsevier	& 1\,136	& 207		& 0.182 \\
C	& RE				& IEEE		& 683		& 73		& 0.107 \\
J	& Softw. Syst. Model.		& Springer	& 549		& 53		& 0.096 \\
J	& Empir. Softw. Eng.		& Springer	& 579		& 25		& 0.043 \\
\hline
\end{tabular}}
\end{table}

\subsection{RQ6: Sufficiency of SE Research Funding}
\label{sec:rq6}
For each engineering branch,
we obtained the total number of publications and PhD dissertations
through the process described in Section~\ref{sec:funding}.
In this way we retrieved a total of
56\,679 software,
28\,514 civil,
23\,969 mechanical,
22\,851 electrical, and
3\,138 chemical engineering publications.
With regard to dissertations,
we identified
55 SE,
302 civil,
2\,018 mechanical,
1\,861 electrical, and
1\,147 chemical engineering records.
Overall,
SE appears first in publications but last in dissertations.

\section{Discussion}
\label{sec:discussion}
More than fifty years after the launch of the first SE conference series,
the SE research discipline can be proud of numerous tangible contributions to practice.
From the survey analysis it appears that
researchers have equipped their industrial partners with a swarm of
new open source software tools,
novel development processes and methods, and
advanced professional practices.
In addition,
they managed to expand the quality and scope of existing proprietary products,
showcasing the importance of maintaining strong academia-industry ties.
However,
the quantitative results of the patent analysis contradict
the self-assessed survey results,
since only a limited number of the collected patents cite any SE paper
of the examined venues.
Still,
almost half (45\%) of the patents citing 69\% of the SE papers have been renewed,
paying out substantial corresponding fees.
(The maintenance fees in the US alone range from \$500
to \$7\,700.\footnote{\url{https://www.uspto.gov/learning-and-resources/fees-and-payment/uspto-fee-schedule\#Patent\%20Maintenance\%20Fee}})
These facts demonstrate the influence of SE papers on valuable patents.

\textbf{Introspective SE Research}
Looking at the practical impact of SE research types, areas, and methods
(Section~\ref{sec:rq1}),
we deduce that in a narrow sense the SE field is introspective in its nature.
Researchers mostly generate knowledge, tools, and reviews
about SE practices and methods to address their own needs~\cite{Sch19}.
This, by definition, limits the potential footprint of SE research
only within its own boundaries,
which is typically not the case for other fields such as
artificial intelligence and machine learning~\cite{XSPD20}.
Consider, for example, the application of the latter in
the petroleum industry~\cite{RP18},
biotechnology~\cite{Oli19},
Internet of Things~\cite{Tur19},
public administration~\cite{Kou17},
healthcare~\cite{YBK18}, and
earthquake engineering~\cite{XSPD20}.
As a result,
we would expect SE research to reach a narrower industrial audience
compared to other science and engineering disciplines.

Yet, in a wider sense SE,
by taming the complexity of software development,
has allowed the exponential growth~\cite{HSG17}
of the sophisticated software intensive systems that underpin
the modern economy, science, and way of living.
This still leaves open the question of the role of SE \emph{research}
in this progress.

\textbf{Cross-disciplinary SE Areas}
SE practice is impacted by cross-disciplinary SE areas.
This is reasonable considering that,
according to the Guide to the Software Engineering Body of Knowledge (SWEBOK)~\cite{swebok},
SE intersects with diverse computer science areas and other disciplines.
A considerable portion of impactful SE research pertains to areas
related to programming languages, compilers, and management (Section~\ref{sec:rq1}).
Furthermore,
looking at the CPC categories of SE patents citing SE research (Section~\ref{sec:cpc})
we observe that some of them
(G06F8/31, 
G06F8/37, 
G06F8/41, 
G06F8/53) 
are associated with the aforementioned areas.
Previous work has also demonstrated that
SE researchers are very interested in human factors~\cite{PA21}.
Consequently,
it might be worth investigating in isolation the impact of
cross-disciplinary and specialized SE areas to SE practice
by studying more specialized venues, such as
the Conference on Programming Language Design and Implementation, 
the International Conference on Object-Oriented Programming Systems, Languages, and Applications, 
the International Conference on Functional Programming, and 
the Symposium on Principles of Programming Languages. 

\textbf{SE Venues}
With regard to SE venues,
one might wonder whether practitioner-oriented ones are more impactful.
We assessed the top SE venues by computing their impact factors
on the basis of SE patent citations (Table~\ref{tab:venues}).
We observe that practitioner-oriented venues,
such as \emph{Software: Practice and Experience} and \emph{IEEE Software},
are higher in the ranking compared to researcher-oriented ones,
such as \emph{Empirical Software Engineering}.
Moreover,
we see a large number of conferences ranked considerably high.
This could be an effect of the practitioner-oriented tracks
of these conferences including
ICSE's \emph{Software Engineering in Practice},
MSR's \emph{Mining Challenge},
ISSTA, FSE and ICSME's \emph{Tool Demonstrations}, and
FSE and ICSME's \emph{Industry Showcase}~\cite{Vya13}.
To increase their industrial appeal,
less practitioner-oriented journals and conferences might want to consider
including dedicated industrial topics, calls, and tracks for contributions,
similar to the aforementioned examples.

\textbf{Correlation Analysis}
From the patent analysis
we obtained no concrete evidence of relationship between patent citations and
impact of SE research in practice.
Although previous work has demonstrated positive correlation
between patent and academic citations~\cite{AH02},
our correlation analysis (Section~\ref{sec:rq2}) did not show
a sufficient association between the two.
In addition,
looking at the most-cited SE papers by SE patents (Table~\ref{tab:top-cited})
we infer that many (\eg\cite{Dea07,Boe84}) are knowledge--
rather than solution-seeking~\cite{SGJ16}.
Their main use as citations in patent documents is most likely the provision of
background information.
Solution-seeking studies usually produce algorithms, models, and tools
to cope with \emph{practical problems}~\cite{Wie09},
whereas knowledge-seeking ones employ cross-disciplinary research methods
(\eg case studies, surveys) to explain \emph{knowledge problems}~\cite{Wie09} of SE practice
(\eg to evaluate and compare different tools,
or to study developers' collaboration)~\cite{SF18}.

\textbf{Knowledge-seeking SE Research}
The strong appearance of knowledge-seeking studies in patent citations could be justified
by the fact that SE (beyond the study scope---Section~\ref{sec:introduction})
is a complex discipline consisting of various dimensions
(also \emph{knowledge areas} in SWEBOK~\cite{swebok}).
Apart from product-related areas (\eg Software Construction and Testing),
there are also areas associated with
the processes, methods, and models employed in the software construction
(SE Process, SE Models and Methods),
the management and cooperation of the development teams (SE Professional Practice), and
the project management of the software development (SE Economics).
These topics are highly relevant to the industry~\cite{DevEco21},
and this could be the reason for their high citation numbers.
Therefore,
the interpretation of the association between patent citations and practical impact
depends on the impact definition.

\textbf{Impact Definition}
The dependence of our analysis on the impact definition,
along with the dissent of some survey respondents from our perceived sense of impact,
constitutes a need for a formal term description.
From the answers to RQ3 and RQ4 (Sections~\ref{sec:rq3},~\ref{sec:rq4})
we infer that the influence of subsequent research was rated both as practical
and non-practical impact by respondents.
To allow future studies building on our work
to quantify and assess the impact of SE research in practice,
academia and industry need to jointly agree on what constitutes impact
based on empirically validated research.
In this regard,
we propose as term description our impact definition in Section~\ref{sec:introduction}.

\textbf{SE Research Funding}
One prominent hindrance to the development and commercialization
of academic SE research products appears to be insufficient funding
(Section~\ref{sec:rq3}).
Although some research ideas may be promising for the industry,
they seem to struggle to evolve due to a lack of financial resources.
These deficiencies concern both project-based and long-term funding,
which is required for a software product's maintenance
after the project completion~\cite{HSM18}.
In addition,
trending industrial research topics,
such as the metaverse, self-driving cars, space, robotics,
and quantum computing~\cite{Ec22},
are often financially unbearable for academia.
In 2021, the R\&D budget of America's top five tech companies
(Amazon, Alphabet, Meta, Apple, Microsoft)
added up to 149 billion dollars,
almost a quarter of America's 2020 public and private R\&D investment
(713 billion)~\cite{Ec22}.
To this end,
SE researchers have repeatedly stressed the need for higher financial support
by society
to be able to perform more realistic studies for the industry
(\eg~\cite{SAAD02,KD04,SDJ07}).
Particularly,
Sj\o{}berg \etal~\cite{SDJ07} argued that,
given the recognized value of software in business~\cite{Boo01},
the discipline should not fall back in funding compared to other fields,
including natural sciences and medicine.
But there is also a matter of priorities resulting from the finite amount
of available resources:
research is the steering wheel of innovation,
but it also needs to cater to financial and societal needs~\cite{BS14}.

Due to the lack of empirical data supporting the aforementioned remarks,
further analysis is needed to assess the sufficiency of SE funding.
To set the path,
we approximated funding based on the number of existing SE publications and PhD dissertations,
and compared them to those of the main engineering branches (Section~\ref{sec:rq6}).
The resulting data are inconclusive:
SE seems to outnumber the other branches in publications,
while trailing in dissertations.
This can be an effect of the employed search method,
leading to false positives or missed records.
Or it could be the case that SE researchers can get away with publishing studies
of low-hanging fruit,
with carefully developed methods and impeccable text,
but having limited practical impact---a practice that may not be widely tolerated
in more mature engineering disciplines.
A comprehensive study of SE research funding could shed light on this issue,
and provide well-grounded insights regarding the funding's adequacy.

\textbf{Tech Transfer Challenges}
Beyond funding,
a set of technology transfer challenges arise
from the related work (Section~\ref{sec:related}) and
the survey answers to RQ3 (Section~\ref{sec:rq3}).
These can be categorized into two broad areas:
industrial obstacles (\ie adoption barriers by practice) and
research challenges (\ie limitations of the proposed work).
Concerning the first,
we observe disregard and questioning of research progress by practice.
In a recent (non peer-reviewed) study Koziolek reports that
practitioners appear willing to participate in case studies and
experiment with new methods from academia,
regardless of the return-on-investment~\cite{Koz22}.
However,
according to two recent large-scale developer surveys
by JetBrains~\cite{DevEco21} and Stack Overflow~\cite{SO21},
the majority of most-used software development tools in practice are developed
in the industry or by practitioners by a vast margin,
and any research results they appear to incorporate were developed
many decades ago~\cite{Roc75,Tic82,RJLK78,Str83}.
The reasons for the industrial provenance of tools used by practitioners could be
the products' technology readiness~\cite{Hed17},
stability, usability, trustworthiness, feature customization,
exclusive customer support and training, and
collaboration ties between companies and software vendors~\cite{SBJ15,SS17}.
Furthermore, companies developing their own software products may be reluctant
to work with open source software coming outside their organization---also known
as the \emph{not invented here syndrome}~\cite{PD15}.
These factors hinder the adoption of academic technology by practice.

Regarding research challenges,
we notice an unfavorable cost-benefit trade-off of some proposed solutions,
despite their high citation numbers and awards.
We summarize the following cases:
useful ideas that are technology immature~\cite{Hed17},
offer conceptual software design solutions~\cite{EU09}, and
require a long time span and re-engineering
to apply in the industry~\cite{ELHC05};
redundant ideas that do not meet industrial needs or quality standards~\cite{LNZ15};
research methods suffering from confirmation bias and unreported implications~\cite{Koz22};
and promising ideas documented in long, inscrutable papers
that require hours of reading~\cite{BORB13}.
Reasons leading to these deficiencies could be
the \emph{publish or perish} culture~\cite{Nat10}
``forcing scientists to produce \emph{publishable} results at all costs''~\cite{Fan10},
excessive competition over collaboration among researchers~\cite{BHW12}, and
reductionist thinking hindering scaling of complex software systems~\cite{RAW12,ESSD08}.

\textbf{Best Practices}
To address the outlined challenges,
several best practices have been proposed in the related literature,
aiming to bridge the gap between academia and industry
(\eg~\cite{GPO16,GPFF19,OCEK00}).
Notable recommendations from Section~\ref{sec:reviews} include:
grounding research on real-world problems and reducing generalizability issues;
publishing more accessible content (\eg through blogs or videos) and
advertising it through social media;
producing more actionable empirical studies, for example, studies on
development productivity and effort estimation techniques;
ensuring open-access availability of research studies;
conducting pilot laboratory tests before industrial releases; and
organizing regular workshops with the industry.
In addition,
adopting the open science and ACM Artifact Badging approaches
can facilitate the adoption and application of research products
by independent, unguided users~\cite{Koz22}.
Collaboration between academia and software vendors,
who already have an established business model,
could also ease the integration of research products into existing tools
through plug-ins~\cite{Koz22}.
From a company perspective,
practitioners should systematically evaluate and improve methods
they have not produced themselves~\cite{Koz22}.
It might be worth investigating empirically the extent to which
such recommended practices have been adopted by industry and academia as well as
any factors impeding their application.

\section{Limitations}
\label{sec:limitations}
Here we present the risks resulting from the patent and survey analyses.
The survey was designed with the stated goal of
examining how SE research has impacted practice.
For this purpose,
we followed recommended guidelines for survey research~\cite{PK01,KP03}.
Response options of two out of three multiple choice questions were adapted
from established literature~\cite{ESSD08,Rou12},
and the complete questionnaire was validated
through two pilot runs.
Although the survey was mainly characterized
as \oecode{interesting} and potentially \oecode{impactful},
its results are limited by the reasons detailed below.

\par{\textbf{Internal Validity}}
Some risks to the internal validity of the study stem from the manual processes
involving subjective judgment.
These include
the identification of the SE-related CPC categories (Section~\ref{sec:cpc}),
the mapping of patent maintenance fee codes to their fee values
(Section~\ref{sec:patents-evaluation}), and
the manual coding of the open-ended survey responses.
Although biases related to human judgment cannot be completely eradicated~\cite{PVK15},
we aimed to reduce this threat by employing established methods~\cite{BKBT07,CS90}.
Manual coding also entails loss of accuracy of responses,
due to the extreme level of their categorization;
we addressed this by assigning multiple (rather than only one) codes to each answer.

Risks also rise from some automated processes we employed.
The identification of SE papers cited by SE patents
through DOI crosschecking, and title and author mapping is not completely accurate,
according to our evaluation (Section~\ref{sec:patents-citations}).
The identification of surveyed practitioners (Section~\ref{sec:survey}) does not account
for 118 (20\%) papers lacking DOIs.
Similarly,
the identification of SE venues in patent citations based on
their full names, abbreviations, and acronyms (Section~\ref{sec:venues}) may have led
to some false positives in Table~\ref{tab:venues}.
In the same Table,
the number of MSR papers does not include records for the years 2010, 2011, and 2018,
because these are not tracked by Scopus.
Publication and PhD dissertation counts for the engineering disciplines
may also include false positives, as confirmed in Section~\ref{sec:discussion}.
Also,
the research funding evaluation (Section~\ref{sec:rq6}) does not make a distinction
between academic and industrial research funding.
One might expect the former to be less impactful than the latter.
Further analysis is required to assess this difference.

A few limitations are associated with the analysis of SE patent citations of SE papers
(Section~\ref{sec:patents-citations}).
The analysis does not account for instances where
an author of a patent may also be involved in the cited publication
either as a direct author or as a collaborator of the research team.
There may also be cases where authors employ both publication means
(scientific publication and patent) and one cites the other
to increase the publication's value.
Moreover,
patent citations were not screened to determine the actual use of SE research.
Consequently, citations that may reference a study as
an example, comparison, statement source, or related work~\cite{KKDS20}
may have skewed the associated quantitative results
in Sections~\ref{sec:rq1},~\ref{sec:rq2}, and
discussion comments (Section~\ref{sec:discussion}).
Although we evaluated patents based on their litigation and maintenance fee events
(Section~\ref{sec:rq2}),
this does not guarantee that all patented products are impactful.
These cases require special treatment to verify the practical impact.

Further concerns may result from the patent-to-science citation linkages
regarding novelty.
Novelty can be a deal breaker when preparing a patent application:
turning an already published idea into a patent is usually very hard,
in many jurisdictions a showstopper~\cite{Moh01}.
However, to cope with this challenge, it is a recommended practice
to adopt the \emph{patent first, publish later} approach~\cite{Moh01},
according to which researchers file a patent application before they publish it
in a research paper.
This approach would result in patent-publication chains, where, for example,
\emph{patent A} is cited by \emph{related-publication A},
\emph{publication A} is cited by \emph{improvement-patent B},
\emph{patent B} is cited by \emph{related-publication B}, and so forth.
The publication-to-patent part of this chain
(\ie \emph{publication A is cited by patent B})
is the center of our patent citation analysis (Section~\ref{sec:rq2}).
As a result,
the cited publication and the citing patent can refer
to different---yet possibly similar---research ideas.
Still, there may also be cases of citations diverging from this approach.

The survey questionnaire is associated with two inherent biases.
Social desirability bias~\cite{Fur86}
(\ie a respondent's potential tendency to appear in a positive light,
for example, by showing they are fair or rational)
is a risk associated with the survey responses
to RQ3 and RQ4.
For instance,
one should not over-interpret that the majority (69\%) of respondents
consider their surveyed work to have impacted either partially or widely SE practice.
To mitigate this issue,
participants were informed that their responses would be published anonymized.
Furthermore,
question-order effect~\cite{Sig81}
(\eg one question may have provided context for the subsequent one)
may have influenced respondents' answers.
Although this effect could have been reduced, for example, by shuffling questions,
we opted to order them in a rational sequence for participants
to recall and comprehend the context of the asked questions.

The collected survey feedback also revealed the following limitations.
The definition of practical impact that was provided to candidates was an earlier version
of the one introduced in Section~\ref{sec:introduction}.
This did not successfully clarify that software development tools
impacted by research should only be industrial,
hence answers to RQ3 and RQ4
may also concern non-industrial tools.
In addition,
the \oecode{subjective} and \oecode{limited definition} of the term
along with its \oecode{quantification process}
roused concerns to some respondents who emphasized that impact
\response{can almost only be assessed by people working
in closely related research areas} \rid{R80}.
The classification process using the 2012 ACM CCS
was deemed \oecode{difficult} and \oecode{irrelevant} by some participants.
Others advised against using institutional e-mail accounts in surveys,
because these are often overloaded with unsolicited messages,
and may thus limit response rate.
To address this, in our manual search
we looked for active e-mail addresses in researchers' personal web pages
(Section~\ref{sec:survey}).
Finally,
the \oecode{author selection process} was not sufficiently documented
in the invitational mail causing confusion to some respondents.

\par{\textbf{External Validity}}
Generalizability concerns arise from
the survey sample selection process (Section~\ref{sec:survey}),
which is limited to the first authors and
the venues of ICSE, TSE, TOSEM, and EMSE,
only includes the most-cited, awarded distinguished, and most influential works,
and involves a small number of practitioners (10\%).
Still, these practitioners are affiliated with 36 diverse companies.
To counterbalance this shortcoming,
we aimed to cover all publication years,
motivated survey candidates to include and review additional impactful publications of theirs,
and also invited them to list any notable papers of other researchers
that may have come to their attention.
Although these concerns prevent us from generalizing the survey findings,
meaningful insights emerged,
which could be amplified through study replication in other research outlets
and author sets.
Furthermore,
the computation of the patent-based impact factors of the top SE venues (Section~\ref{sec:venues})
is affected by the restricted year range (2009--2019).
As explained,
the rationale was to approximate publication counts of all years,
preventing any risk that could occur from different venue start dates.

\section{Concluding Remarks}
\label{sec:conclusions}
We investigated the impact of SE research in practice
through a systematic analysis of science linkages between SE research and SE patents,
and a survey on authors of top-notch publications.
Specifically,
we identified impactful types, areas, and methods of SE research,
the outcomes of impactful research, and
its main practical impacts on information technology, society, and industry.
We further assessed the impact of SE venues and
the sufficiency of SE research funding by comparing it to the main engineering branches'.
To address these,
we collected 11\,419 papers from ICSE, TSE, TOSEM, and EMSE between 1975--2017,
and complemented them with their assigned topics, citation counts, and awards.
We also retrieved a set of 304\,368 SE patents,
and found 1\,690 papers cited by 4\,354 of them.
To assess patents' value,
we analyzed their litigation cases and maintenance fee events.
Through a survey on 475 authors
of 593 top-cited and awarded papers
(26\% response rate),
we complemented our study results with quantitative and qualitative insights.
For the venues' impact
we computed their patent-based impact factors, while
for the adequacy of SE research funding
we retrieved the research publication and dissertation counts of SE
and the main engineering branches.
The study's key findings are summarized below.
\begin{itemize}[leftmargin=*]
\item SE researchers have equipped practitioners with various tools, processes,
and methods, and improved many existing solutions.
Moreover, practitioners seem to value knowledge-seeking studies.
\item SE practice is impacted by cross-disciplinary SE areas,
hence it could be of value to assess this influence by studying in more depth
some specialized venues.
\item Practitioner-oriented tracks in conferences may enhance their impact.
A dedicated study of these tracks could provide more insights as well as
useful recommendations to organization and program committees.
\item Academia and industry could jointly agree on a formal impact term description
based on empirically validated research and
backed by key performance indicators
to set a common ground for subsequent research on the topic.
\item There is a claim for higher funding in SE research,
which we cannot corroborate through our analysis on engineering dissertations
and publications, or literature search.
A comprehensive empirical study could shed light on the matter.
\end{itemize}

\par{\textbf{Research Agenda}}
There are various directions for extending the investigation of the SE research impact
in practice---some are listed in the preceding paragraphs.
Other ideas include:
studying the patenting behavior by SE research area, and
investigating the practical impact of less patent-focused areas
(\eg agile/lean methods, test automation);
conducting in-depth systematic analyses of SE patent citations
to delve into the use of academic research by patents; and
comparing reports of industry impact
(\eg company statements, software development metrics)
with venue, funding, and citation metrics.
With these final remarks
we aim to steer academia's attention towards some research topics
requiring further investigation, and
begin a discussion on how we, the researchers, can increase our footprint on practice.

\section*{Acknowledgments}
We would like to thank ACM for providing us
with the \emph{ACM DL Abstracts and Titles for Research Purposes} database.
This work has received funding from the
European Union's Horizon 2020 research and innovation programme
under grant agreement No. 825328 (FASTEN project).

\bibliographystyle{IEEEtran}
\bibliography{swimpact}

\end{document}